\documentclass[aip,nofootinbib]{revtex4}
\usepackage[T1]{fontenc}
\usepackage{color}
\usepackage{graphicx}
\usepackage{amsmath,amssymb}
\usepackage{scalefnt,color}

\makeatletter

\newcommand{\order}[1]{{\cal O}(#1)}
\newcommand{\dd}{\mathrm{d}}
\newcommand{\code}{\sf}
\newcounter{rhnotecount}
\newcommand{\op}{\mathcal{O}}
\newcommand{\eqn}[1]{Eq.\,(\ref{#1})}
\newcommand{\eqns}[1]{Eqs.\,(\ref{#1})}
\newcommand{\mtop}{m_t}

\newcommand{\mhiggs}{m_H}
\newcommand{\pt}{p_\mathrm{T}}
\newcommand{\qcd}{{\abbrev QCD}}
\newcommand{\lhc}{{\abbrev LHC}}
\newcommand{\nlo}{{\abbrev NLO}}
\newcommand{\lo}{{\abbrev LO}}
\newcommand{\sm}{{\abbrev SM}}
\newcommand{\ewsb}{{\abbrev EWSB}}
\newcommand{\abbrev}{\scalefont{.9}}

\usepackage[parfill]{parskip}
\usepackage[plain]{fancyref}
\usepackage{varioref}
\usepackage{nicefrac}
\usepackage{flafter}
\usepackage{placeins}

\begin{document}
\title{
\begin{flushright}
\vspace*{-3em}
\small\sf August 2013 --- WUB/13-08
\vspace*{3em}
\end{flushright}
Probing the nature of the Higgs-gluon coupling
}

\author{Robert V. Harlander}
\email{robert.harlander@uni-wuppertal.de}
\author{Tobias Neumann}
\email{tobias.neumann@uni-wuppertal.de}
\affiliation{Bergische Universit\"at Wuppertal, 42097 Wuppertal, Germany\\}

\begin{abstract}
  One and two-jet observables of dimension-7 Higgs-gluon coupling
  operators are studied as probes of possible deviations from the
  top-loop induced gluon-Higgs coupling.  We discuss the case of both a
  scalar as well as a pseudo-scalar Higgs boson and show that higher
  order operators can give visible deviations from \sm{} distribution
  shapes.
\end{abstract}

\maketitle


\section{Introduction}

The mechanism of electroweak symmetry breaking (\ewsb{}) is in its phase
of identification. The recent
discovery\,\cite{Aad:2012tfa,Chatrchyan:2012ufa} of a Higgs boson at the
Large Hadron Collider (\lhc{}) provides a new probe for physics beyond
the Standard Model (\sm{}) through precision physics. Up to now, the
signals are in very good agreement with \sm{} predictions, but there is
still room for traces of an extended theory.

The production and decay rates of the \sm{} Higgs boson have been
predicted by higher order theoretical calculations~(see
Refs.~\cite{Heinemeyer:2013tqa,Dittmaier:2012vm,Dittmaier:2011ti}). The
dominant production mechanism is through gluon fusion, where the
coupling of the gluons to the Higgs boson is mediated predominantly by a
top-quark loop. It should be one of prior goals of the \lhc{} to test
the character of this gluon-Higgs coupling.

The loop-induced coupling differs particularly strongly from a
point-like coupling if an external mass scale of the process becomes
larger than the top-quark mass, and thus allows to resolve the
non-trivial structure of the Higgs-gluon vertex. This suggests to look
at processes involving large transverse momenta of the Higgs boson or an
associated jet, for example, and compare the \sm{} prediction with a
point-like coupling.

A general description of a point-like gluon-Higgs coupling that is
compatible with the gauge symmetries of the \sm{} can be obtained in
terms of an effective theory. The leading operator for a scalar Higgs,
$\op_1 = HF^a_{\mu\nu}F^{a,\mu\nu}$ is of mass dimension 5. In the limit
of infinite top mass, $\mtop\to \infty$, and neglecting the Yukawa
coupling of lighter quarks, the \sm{} gluon-Higgs coupling reduces to
this single operator; for finite $\mtop$, it allows for a reasonable
approximation of the \sm{} gluon-Higgs coupling and is typically used
for evaluating perturbative correction factors to the \sm{} cross
section prediction. Differential cross sections based on $\op_1$ have
therefore been studied quite extensively. The purpose of our paper is to
see the effect of higher dimensional operators on the most important
kinematical distributions. These operators of mass dimension~7 are well
known; they will be recalled in the next section.

The philosophy of our analysis is thus as follows: assume that it turns
out that the gluon-Higgs coupling is {\it not} (or not mainly) induced
by a top-quark loop as in the \sm{}, but by point-like vertices
generated by some new physics at higher scale. The coupling strengths of
the effective operators (at least one of them) has to be \sm{}-like in
order to account for the signal strength observed experimentally. It
will then be important to carefully consider distributions and compare
them with the predictions based on the effective Lagrangian to be
introduced below.  Our discussion includes both the production of scalar
and pseudo-scalar Higgs bosons (both denoted as $H$ in what follows) and
will be based on $H$+1- and $H$+2-jet observables.

Before we begin our discussion, let us first list some complementary
approches to quantify deviations from \sm{} \ewsb{}:
\begin{itemize}
    \item The ``{\abbrev LHCHXSWG} interim recommendations to explore
      the coupling structure of a Higgs-Like particle''
      \cite{LHCHiggsCrossSectionWorkingGroup:2012nn} assume a single \sm{}
      Higgs-like state $H$ at roughly $125\,\mathrm{GeV}$ and look at
      its production and decay through $(\sigma\cdot\text{BR})(ii\to
      H\to ff)=\sigma_{ii}\Gamma_{ff}/\Gamma_{H}$, where $ii$ and $ff$
      name the initial and final state configurations. They take \sm{}
      deviations into account by modifying strengths of SM operators but
      keeping the same tensor structure.
\item The {\abbrev SILH} effective model \cite{Giudice:2007fh} corresponds
  to an effective theory for \sm{} fields plus a Higgs-like particle.
  The leading \qcd{} operator for Higgs production in {\abbrev SILH} is
  just $H\, F_{\mu\nu}^{a}F^{a\,\mu\nu}$, which is also included in our
  analysis, of course.
\item The approach in \cite{Passarino:2012cb} takes {\abbrev BSM}
  physics into account by the most general set of dimension-5 operators
  consistent with symmetries and therefore augments the ``LHCHXSWG
  recommendation''~\cite{LHCHiggsCrossSectionWorkingGroup:2012nn}.  It
  can parametrize less \sm{}-like physics and possibly include {\abbrev
    SILH} physics.
\end{itemize}

Note that none of these approaches takes higher dimensional gluonic
operators into account, which is the subject of this paper. Our
motivation is that the dominant Higgs production mechanism at the \lhc{}
is gluon fusion, and we want to study the sensitivity of kinematical
distributions of the Higgs boson and the associated jets to the precise
form of the Higgs-gluon coupling. For similar, though more
model-specific analyses, see
Refs.\,\cite{%
Hankele:2006ma,Langenegger:2006wu,Campanario:2010mi,Bagnaschi:2011tu},
for example.

The remainder of this paper is organized as follows. In
\fref{sec:basis}, the basis of dimension-5 and -7 operators is recalled
that couple a scalar or a pseudo-scalar particle to gluons in a
gauge invariant way. We also briefly describe some technical issues of
our study. \Fref{sec:distros} presents distributions of $H$+1-jet and
$H$+2-jets observables as induced by the formally leading and
sub-leading terms in the effective theory. The \sm{} case is reproduced
for comparison. In \fref{sec:higher}, we also consider terms that are
formally suppressed by higher powers of the ``new physics scale''
$\Lambda$ which occurs in the effective theory. In \Fref{sec:conclusions}
we present our conclusions.


\section{Basis of dimension-7 operators and their 
implementation}\label{sec:basis}

This section describes the operator basis used in our calculation. We
start with the effective Lagrangian for the coupling of a scalar boson
to gluons, and generalize the discussion to pseudo-scalar bosons in the
subsequent section.


\subsection{Scalar Higgs boson}

The effective Lagrangian involving operators through mass dimension~7
which couple a scalar Higgs boson $H$ to gluons can be written as
\cite{Gracey:2002he,Neill:2009tn} (see also Ref.\,\cite{Germer:diplom})
\begin{equation}
\begin{split}
\mathcal{L}&= \frac{C_1}{\Lambda}\op_1 
+ \sum_{n=2}^5 \frac{C_n}{\Lambda^3}\op_n
\label{eq:leff}
\end{split}
\end{equation}
\begin{equation}
\begin{split}
\op_1&=HF_{\mu\nu}^{a}F^{a\,\mu\nu}\,,\quad
\op_2 = HD_{\alpha}F_{\mu\nu}^{a}D^{\alpha}F^{a\,\mu\nu}\,,\quad
\op_3 =
HF_{\nu}^{a\,\mu}F_{\sigma}^{b\,\nu}F_{\mu}^{c\,\sigma}f^{abc}\,,
\\
\op_4 &= HD^{\alpha}F_{\alpha\nu}^{a}D_{\beta}F^{a\,\beta\nu}\,,\quad
\op_5 = HF_{\alpha\nu}^{a}D^{\nu}D^{\beta}F_{\beta}^{a\,\alpha}\,,
\label{eq:opeff}
\end{split}
\end{equation}
where
\begin{equation}
\begin{split}
    F_{\mu\nu}^a = \partial_\mu A_\nu^a - \partial_\nu A_\mu^a - g_s
    f^{abc} A_\mu^b A_\nu^c\,, \qquad D_\mu A_\nu^a = \partial_\mu
    A_\nu^a - g_s f^{abc} A_\mu^b A_\nu^c\,,
\end{split}
\end{equation}
with the gluon field $A_\mu^a$.  The strong coupling is denoted by
$g_s$, and $f^{abc}$ are the {\abbrev SU(3)} structure constants. We
remark that, for an on-shell Higgs boson, the operators in
\eqn{eq:opeff} are not linearly independent. Instead, one finds
$\mhiggs^2\op_1 = 4\op_5-2\op_2+4g_s\op_3$, where $\mhiggs$ is the Higgs
mass, and thus one of $\op_2$, $\op_3$, and $\op_5$ could be eliminated
from our analysis. Nevertheless, we find the basis in \eqn{eq:opeff}
convenient and therefore stick to this redundancy.

The mass parameter $\Lambda$ in \eqn{eq:leff} is undetermined {\it a
  priori}; in the \sm{}, it is the top-quark mass $\mtop$, for
example. Matching the effective Lagrangian of \eqn{eq:leff} to the \sm{}
allows one to derive perturbative expressions $C_i^\text{\sm}$ for the
Wilson coefficients $C_i$. For example, $C_1^\text{\sm}$ is known
through N$^4$LO\,\cite{Chetyrkin:2005ia,Schroder:2005hy}; explicit
expressions for the $C_n^\text{\sm}$ ($n\in\{2,\ldots, 5\}$), on the
other hand, have been obtained only at \nlo{}\,\cite{Neill:2009tn},
where we give the \lo{} expressions as
follows:\footnote{Ref.\,\cite{Neill:2009tn} misses a factor of
  $-\nicefrac{3}{4}$ in $C_{3}^\text{\sm}$, at least at \lo{}, which we
  correct here.}
\begin{equation}
\begin{split}
C^\text{\sm}_{1} & =
\frac{g_s^{2}\lambda_t}{48\pi^{2}} 
+\mathcal{O}(g_s^{4})\simeq2.2\cdot10^{-3}\,,\\ C^\text{\sm}_{2}
& =
\frac{-7g_s^{2}\lambda_t}{2880\pi^{2}}
+\mathcal{O}(g_s^{4})\simeq-2.6\cdot10^{-4}\,,\\ C^\text{\sm}_{3}
& =
-\frac{g_s^{3}\lambda_t}{240\pi^{2}}
+\mathcal{O}(g_s^{5})\simeq-5.3\cdot10^{-4}\,,\\ C^\text{\sm}_{4}
& =
\frac{g_s^{2}\lambda_t}{1440\pi^{2}}
+\mathcal{O}(g_s^{4})\simeq7.3\cdot10^{-5}\,,\\ C^\text{\sm}_{5}
& =
\frac{g_s^{2}\lambda_t}{80\pi^{2}}
+\mathcal{O}(g_s^{4})\simeq1.3\cdot10^{-3}\,,
\label{eq:c1sm}
\end{split}
\end{equation}
where $\lambda_t = \mtop/v$ is the top-quark Yukawa coupling, and the
values $\mtop=172$\,GeV, $v=246$\,GeV, and $g_s=\sqrt{4\pi\alpha_s}$,
with $\alpha_s=0.118$ have been inserted in order to arrive at a
numerical illustration for the size of these coefficients.

Nominally, contributions of $\op_1$ are suppressed by $1/\Lambda^2$ in
physical quantities, mixed terms of $\op_1$ with $\op_2$ to $\op_5$ are
suppressed by $1/\Lambda^4$, etc. Note, however, that in the \sm{}, the
matching coefficients are proportional to 
$\lambda_t\sim \mtop/v = \Lambda/v$. This cancels the
pre-factor $1/\Lambda$ in \eqn{eq:leff} and thus the overall
suppression. Since we want to keep the discussion as general as
possible, we will mostly ignore the suppression of the higher
dimensional operators; all that is relevant to us is the influence of
the individual operators on the shape of the distributions.

\paragraph{Remark on operators containing quark fields.}
The operators $\op_4$ and $\op_5$ can be rewritten to operators
containing two and one quark bilinears according to the \qcd{} equations
of motion~\cite{Arzt:1993gz,Georgi:1991ch,Politzer:1980me}:
\begin{equation}
\begin{split}
D^{\mu}F_{\mu\nu}^{a}(x)=g_{s}\bar{\Psi}(x)\gamma_{\nu}t^{a}\Psi(x)\,.
\label{eq:eom}
\end{split}
\end{equation}
In case of gluodynamics, these two operators then vanish
\cite{Gracey:2002he}. They only contribute when their corresponding
quark-representation can occur in the given process/Feynman diagrams.


\subsection{Pseudo-Scalar operators}

In addition to the production of a scalar Higgs boson, we also consider
the pseudo-scalar analogue which, for the sake of simplicity, we also
denote by $H$ in this paper.\footnote{It should be clear from the
  context whether $H$ denotes the scalar or the pseudo-scalar Higgs
  boson.} The corresponding effective operators are obtained from the
scalar case by replacing one of the field strength tensors by its dual
$\tilde{F}_{\mu\nu}^{a}=\frac{1}{2}\epsilon_{\mu\nu\rho\sigma}F^{a\,\rho\sigma}$,
with the Levi-Civita symbol $\epsilon_{\mu\nu\rho\sigma}$. We therefore obtain
\begin{equation}
\begin{split}
\mathcal{L}&= \frac{\tilde{C}_1}{\Lambda}\op_1 
+ \sum_{n=2}^5 \frac{\tilde{C}_n}{\Lambda^3}\op_n\,,
\label{eq:lefftilde}
\end{split}
\end{equation}
\begin{equation}
\begin{split}
\tilde\op_1&=H\tilde{F}_{\mu\nu}^{a}F^{a\,\mu\nu}\,,\quad
\tilde\op_2 = HD_{\alpha}\tilde{F}_{\mu\nu}^{a}D^{\alpha}F^{a\,\mu\nu}\,,\quad
\tilde\op_3 =
H\tilde{F}_{\nu}^{a\,\mu}F_{\sigma}^{b\,\nu}F_{\mu}^{c\,\sigma}f^{abc}\,,
\\
\tilde\op_4 &= HD^{\alpha}\tilde{F}_{\alpha\nu}^{a}D_{\beta}F^{a\,\beta\nu}\,,\quad
\tilde\op_5 = H\tilde{F}_{\alpha\nu}^{a}D^{\nu}D^{\beta}F_{\beta}^{a\,\alpha}\,.
\label{eq:opefftilde}
\end{split}
\end{equation}
We generated the Feynman rules for the operators $\op_n$ and
$\tilde\op_n$ ($n=1,\ldots,5$) using {\code
  LanHEP}\,\cite{Semenov:2008jy} and confirmed their validity with
{\code FeynRules}\,\cite{Christensen:2008py}. A non-zero contribution of
the operator $\tilde\op_4$ involves at least six gluons; therefore, it
does not appear in our numerical analysis below. Furthermore, similar to
the scalar case, the remaining operators are not linearly independent
for an on-shell Higgs boson, but for convenience we will include all of
them in our analysis.

With the obtained vertices, the necessary \lo{} Feynman graphs for
$H$+1-jet and $H$+2-jets amplitudes were generated with {\code \abbrev
  DIANA}\,\cite{Tentyukov:1999is} as {\code\abbrev
  FORM}\,\cite{Vermaseren:2000nd} code.  The analytical expressions for the
matrix elements were then calculated by inserting the Feynman rules,
where we used Feynman gauge and Faddeev-Popov ghosts, as well as an
axial gauge for cross checking. The calculation of the cross sections
was performed by means of standard {\code\abbrev VEGAS} integration.

The operator $\tilde\op_4$ only contributes at higher order or at 3-jet
production. This can easily be seen by switching to the quark
representation using \qcd{} equations of motion.  The only non-vanishing
Feynman-vertex contribution contains three gluons and a quark-antiquark
pair, thus it does not appear in the quantities considered in this paper
at \lo{}.  All lower order vertices vanish due to the contraction of
equal-momentum vectors with the Levi-Civita symbol.



\section{Kinematical distributions}\label{sec:distros}


In what follows, we consider the transverse momentum
distribution of the Higgs-boson when it is produced in association with
one or two jets, $H$+1-jet and $H$+2-jet for short. In the $H$+1-jet
case, we compare the distributions obtained for the point-like vertices
$\op_n$ and $\tilde\op_n$ ($n=1,\ldots,5$) with the \sm{}-like
loop-induced coupling.

The differential cross section based on the Lagrangian of \eqn{eq:leff}
can be written as
\begin{equation}
\begin{split}
\dd\sigma = \sum_{i,j=1}^5\dd\sigma_{ij}\,,
\end{split}
\end{equation}
where $\dd\sigma_{ij}$ is due to terms of the form $\op_i\op_j^\dagger$.
In order to be independent of the actual size of the Wilson
coefficients, we consider kinematical distributions normalized to their
respective contribution to the inclusive cross section $\sigma_{ij}$,
i.e., $\dd\sigma_{ij}/\sigma_{ij}$. Absolute effects on the
distributions within a given model, i.e., for concrete values of the
Wilson coefficients $C_i$ and the ``new physics scale'' $\Lambda$, can
be derived by combining these normalized distributions with the
numerical values for the total cross sections $\sigma_{ij}$ provided in
Appendix\,\ref{app:norm}.  Notice that interference terms with $i\neq j$
need not be positive definite.

We will show distributions for \lhc{} proton-proton collisions, split
into the partonic gluon-gluon ($gg$), the gluon-quark ($qg$), and
quark-quark ($qq$) initial states ($qq$ includes quark-antiquark as well
as different-flavor initial states).  At \lo{}, the operator $\op_5$
contributes only to the $gq$-channel because, according to \eqn{eq:eom},
it can be rewritten to include a quark bilinear term. Analogously,
$\op_4$ can be rewritten to contain two quark bilinears, and therefore
only contributes to the $qq$ channel.  For the $gq$-channel, $\op_3$
does not contribute because its Feynman rules involve at least three
gluons.

All distributions are generated for a Higgs mass of
$\mhiggs=125\,\mathrm{GeV}$ and a center of mass energy of
$\sqrt{s}=13\,\mathrm{TeV}$. For the $H$+1-jet $\pt$-distributions, the
choice of factorization and renormalization scale is
$\mu=\sqrt{\mhiggs^2 + \pt^2}$ and in case of $H$+2-jet cross sections
the jet-$\pt$ geometric mean $\mu=\sqrt{ \pt(j_1)\, \pt(j_2) }$. Due to
our normalization, the results are largely unaffected by changes of
$\mhiggs$ and $\sqrt{s}$.


\subsection{$H$+1-jet cross sections}

\begin{figure}
\includegraphics[width=\textwidth]{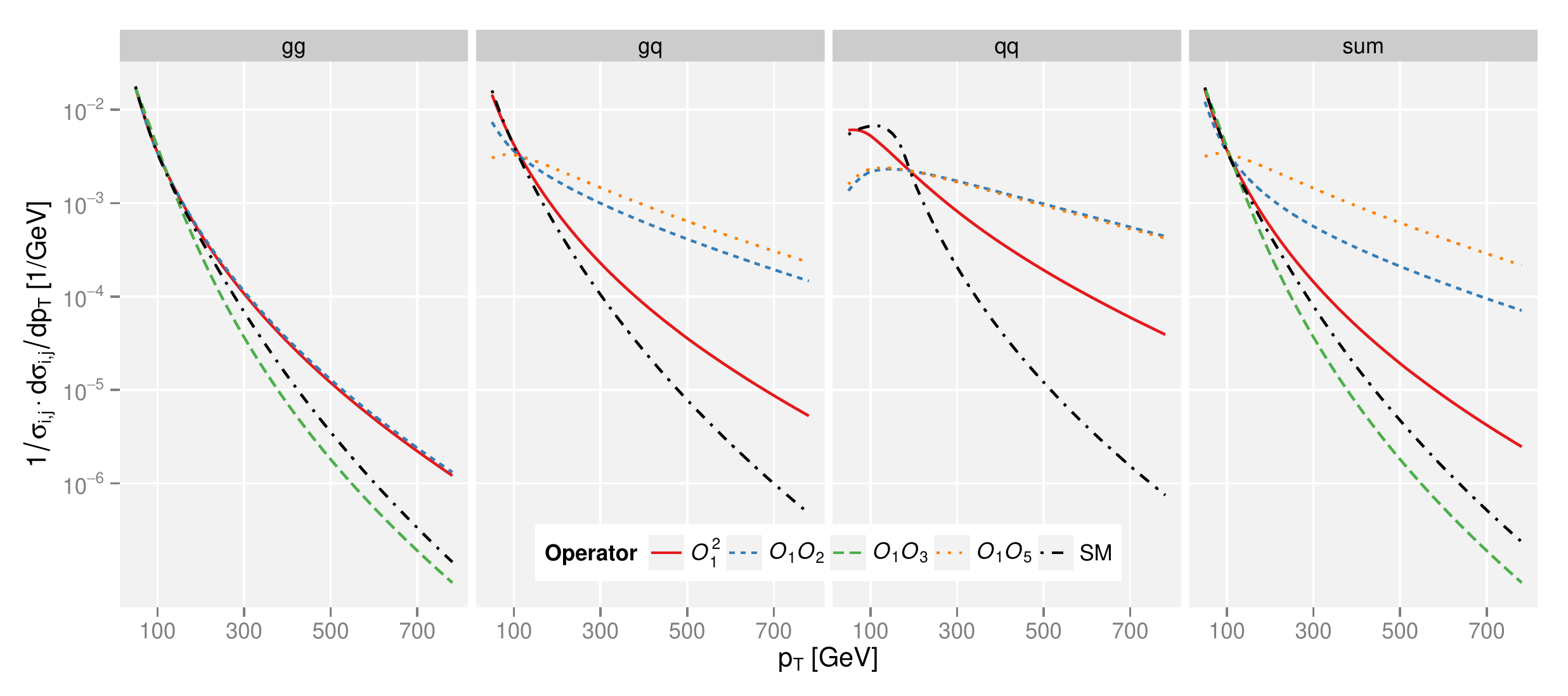}
\caption{\label{fig:ptdist-LO-suppression}Normalized Higgs transverse
  momentum distributions for scalar coupling operators.
  The normalization factors $\sigma_{ij}$ are given in \fref{tab:sigmaijpt}.}
\end{figure}

\begin{figure}
\includegraphics[width=\textwidth]{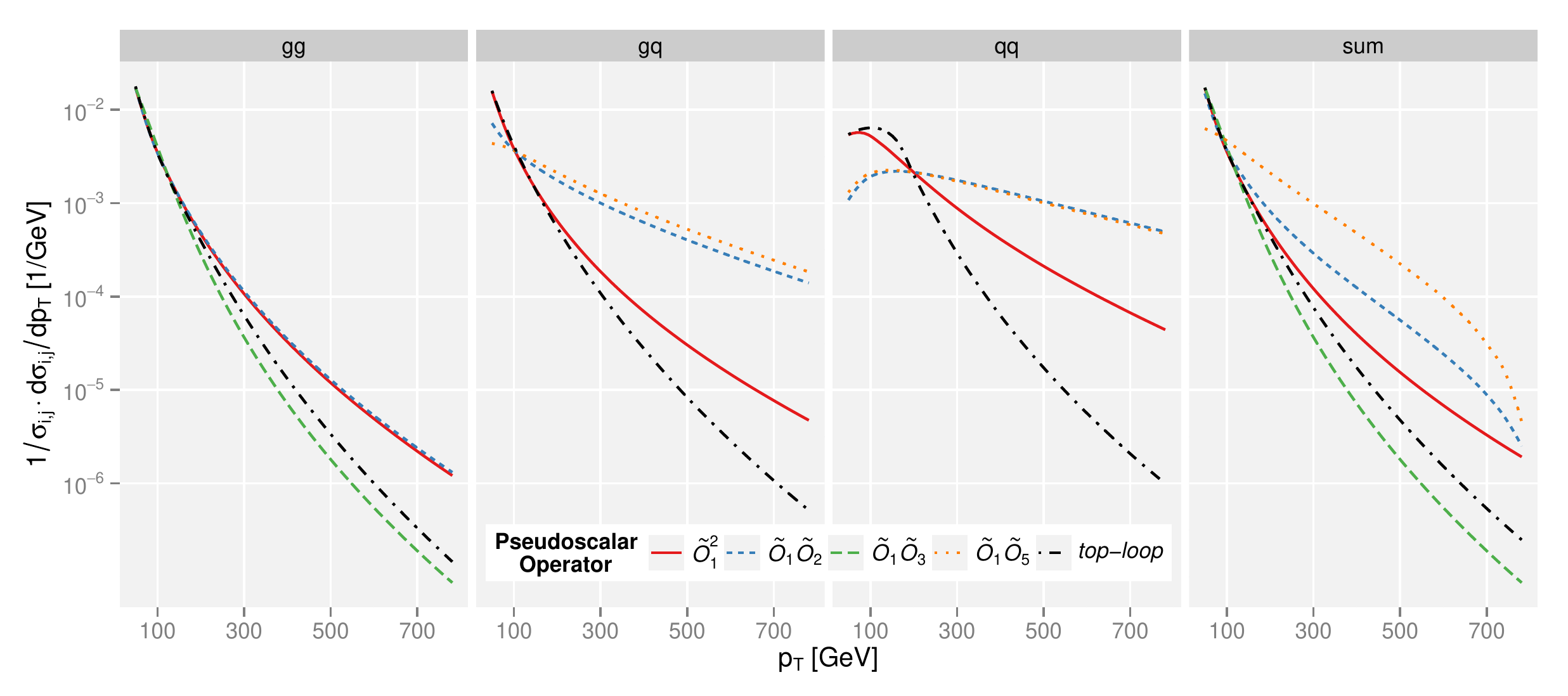}
\caption[]{\label{fig:ptdist_pseudo-LO-suppression} Same as
  \fref{fig:ptdist-LO-suppression}, but for pseudo-scalar coupling
  operators.  Note that the $gg$ channel is identical to the scalar
  case.  The $\sigma_{ij}$ are given in
  \fref{tab:sigmaijpttilde}.}
\end{figure}

First we consider the Higgs transverse momentum ($\pt$) distribution in
$H$+1-jet production for scalar and pseudo-scalar Higgs bosons in
\fref{fig:ptdist-LO-suppression}
and~\ref{fig:ptdist_pseudo-LO-suppression}.  In both cases, one observes
large differences in the $\pt$ shape of the individual terms. For
comparison, we show the curve which corresponds to \sm{}-like Higgs
production through a top-loop, obtained from the program {\tt
  SusHi}\,\cite{Harlander:2012pb} (curves denoted ``\sm{}'' for the
scalar and ``top-loop'' for the pseudo-scalar). Note, however, that
strictly speaking the latter receives an additional $\pt$ dependence
through its proportionality to $\alpha_s^2(\mu)$, with
$\mu=\sqrt{\mhiggs^2+\pt^2}$. In order to properly compare to the
predictions from the effective theory, we have divided the \sm{} and
top-loop distribution by this factor.

The panel named ``sum'' in \fref{fig:ptdist-LO-suppression} and
\ref{fig:ptdist_pseudo-LO-suppression} shows the sum over the partonic
channels for fixed $ij$.  The respective inclusive cross sections to
which these curves are normalized are listed in \fref{tab:sigmaijpt} and
\fref{tab:sigmaijpttilde}.

\begin{figure}
    \includegraphics[width=\textwidth]{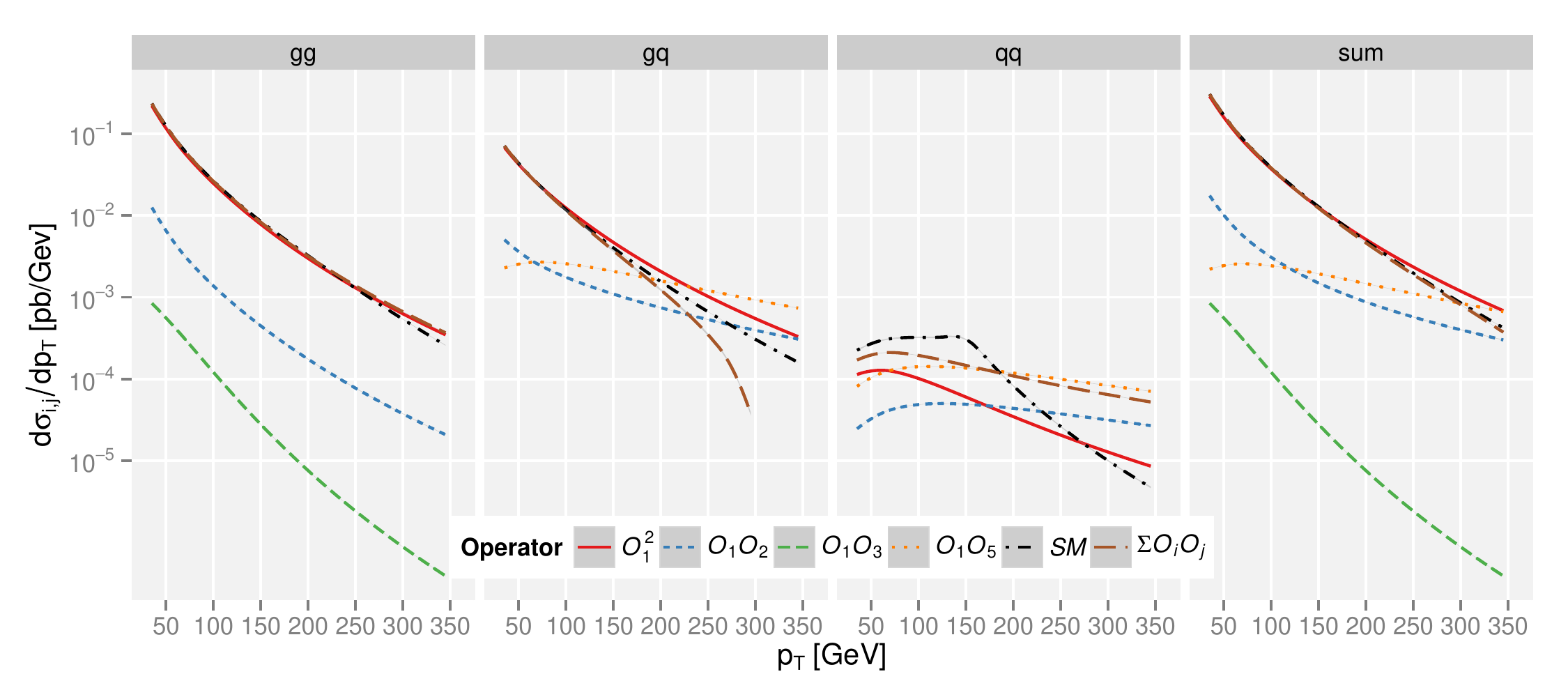}
    \caption[]{\label{fig:sm-coefficients}
      Higgs transverse momentum distributions with \sm{} matching
      coefficients, resulting in a total $\mtop^2$-suppression with
      respect to $C_1^2$. Please note that in case of the $gq$ and
      summed channel the crossterm $\op_1\op_5$ has been multiplied with
      $-1$. For the $qq$ channel the crossterm $\op_1\op_2$ has been multiplied
      with $-1$.}
\end{figure}

As a check, we used these results and combined them with the \sm{}
Wilson coefficients of \eqn{eq:c1sm} in order to reproduce the first two
non-vanishing terms in the $1/\mtop$ expansion for the
$\pt$-distribution. The results are shown in
\fref{fig:sm-coefficients}. They give some deeper insight into the
observations of Ref.\,\cite{Harlander:2012hf}. For the $gg$ channel, the
interference terms of $\op_1$ with the higher order operators have a
very similar shape as the dominant $\op_1\op_1^\dagger$
contribution. The effective theory approach to Higgs production in the
\sm{} therefore works extremely well for the $gg$ subchannel, as
observed also at \nlo{} \qcd{} in Ref.\,\cite{Harlander:2012hf}. For the
$qg$ channel, on the other hand, the various contributions differ rather
strongly among each other. The $\op_1\op_5^\dagger$ term is even
negative; since its magnitude hardly decreases towards larger $\pt$, it
drives the $qg$ channel to negative values. But note that, in contrast
to \fref{fig:ptdist-LO-suppression}, the scale $\Lambda$ enters these
results, and that it is set to $\Lambda=\mtop$; the higher dimensional
operators therefore become numerically dominant at $\pt\gtrsim\mtop$.

In contrast to the $\pt$ distribution,
we do not observe any significant differences for the rapidity
distributions among the operators, which is why we refrain from showing
these results here.


\subsection{Two jet cross sections}

For two jets there are considerably more interesting observables than
for the 1-jet case. In particular the azimuthal angle difference
$\Delta\Phi_{jj}$ between the two jets and the rapidity separation
$\Delta\eta_{jj}$ are well known 2-jet observables for gluon fusion and
weak boson fusion ({\abbrev
  WBF})~\cite{DelDuca:2003ba,DelDuca:2001fn,Rainwater:1997dg,Klamke:2007cu}.
For example, suitable cuts in $\Delta\Phi_{jj}$\,distributions allow one
to discriminate a scalar from a pseudo-scalar Higgs boson in $H$+2-jets
production\,\cite{Hankele:2006ja}. Also, $\Delta\Phi_{jj}$ and
$\Delta\eta_{jj}$ have been proposed to distinguish $H$+2-jet production
through gluon fusion from {\abbrev WBF}\,\cite{DelDuca:2001fn}. We will
study these distributions for the higher dimensional operators of
\fref{sec:basis} and see how they may affect the conclusions
drawn from previous studies.

In the following we will use ``inclusive'' cuts for the
$\Delta\eta_{jj}$-distribution,
\begin{equation}
\begin{split}
p_{j\perp}>20\,\mathrm{GeV}\,,\quad\left|\eta_{j}\right|<5\,,\quad
R_{jj}>0.6\,,
\label{eq:inccuts}
\end{split}
\end{equation}
where $R_{jj}=\sqrt{(\Delta\eta_{j})^{2}+(\Delta\Phi_{jj})^{2}}$, and
additional ``{\abbrev WBF} cuts''
\begin{equation}
\begin{split}
\Delta\eta_{jj}=\left|\eta_{j1}-\eta_{j2}\right|>4.2\,,\quad
\eta_{j1}\cdot\eta_{j2}<0\,,\quad
m_{jj}>600\,\mathrm{GeV}
\label{eq:wbfcuts}
\end{split}
\end{equation}
for the $\Delta\Phi_{jj}$ distribution, where $m_{jj}$ is the invariant
mass of the two jets.

\begin{figure}
\includegraphics[width=\textwidth]{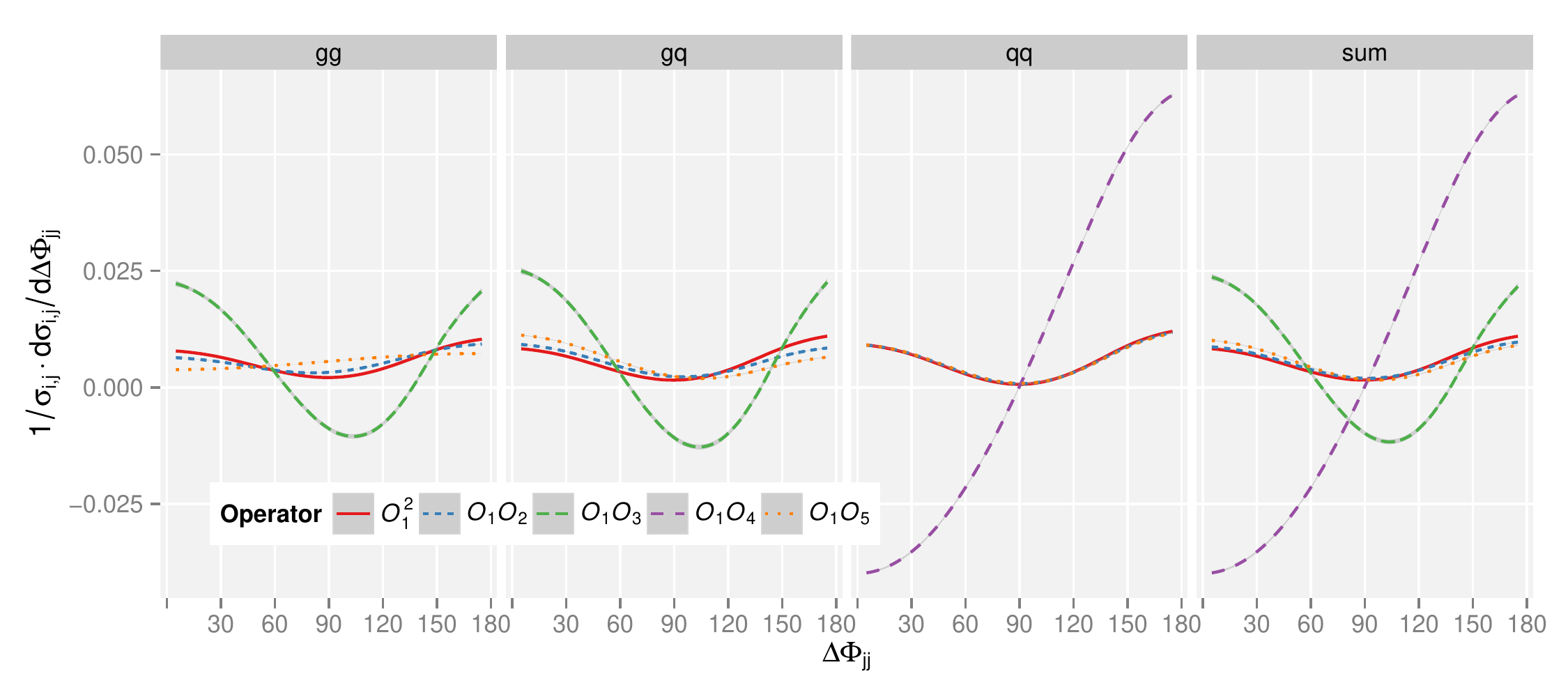}
\caption{Normalized distributions for azimuthal angle difference of the
  two final state jets for scalar operators.  The $\sigma_{ij}$ are
  given in  \fref{tab:phiij}.}
\label{fig:phidist-LO-suppression}
\end{figure}

\begin{figure}
\includegraphics[width=\textwidth]{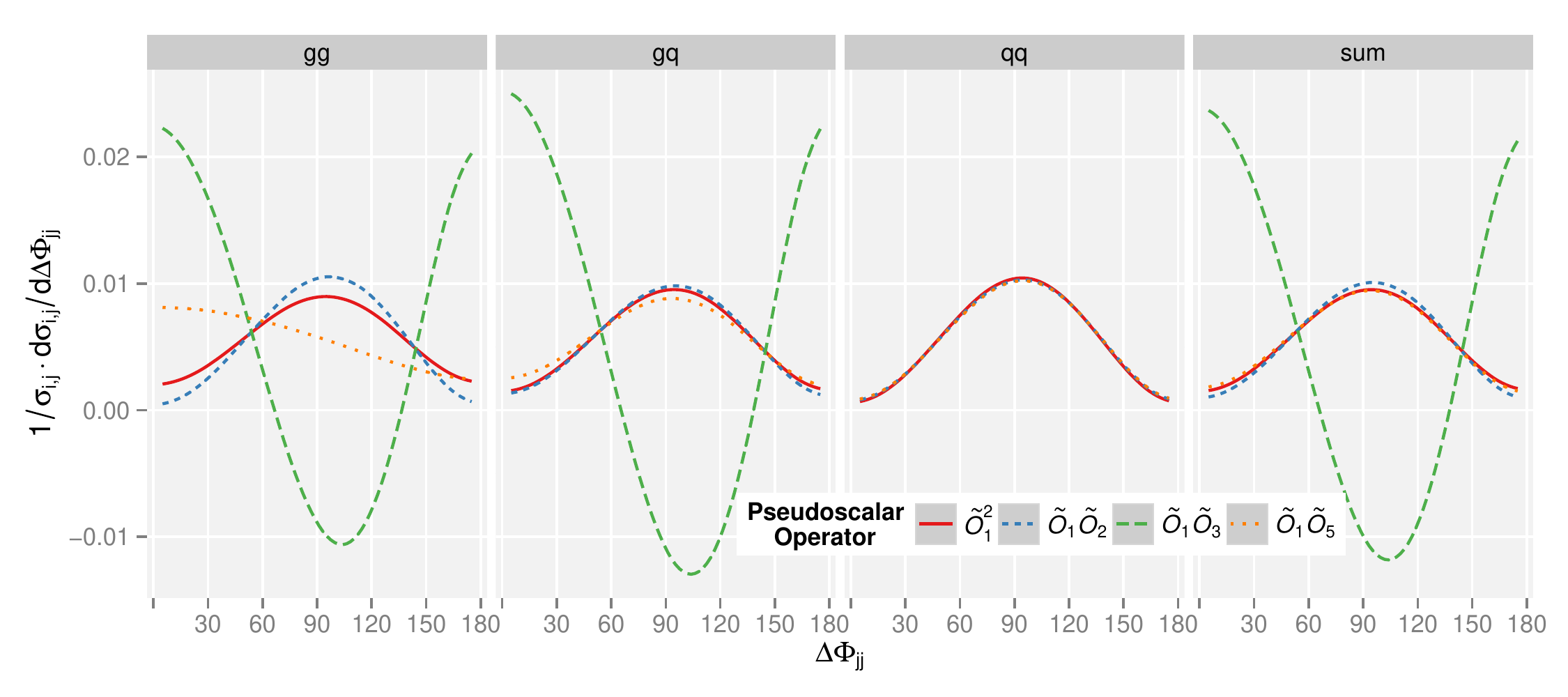}
\caption{\label{fig:phidist-pseudo-LO-suppression}Same as
  \fref{fig:phidist-LO-suppression}, but for pseudo-scalar
  operators. The $\sigma_{ij}$ are given in \fref{tab:phiijtilde}.}
\end{figure}

The $\Delta\Phi_{jj}$ distributions are shown in
\fref{fig:phidist-LO-suppression} and
\fref{fig:phidist-pseudo-LO-suppression}. The red curve, corresponding
to the contribution from $\op_1\op_1^\dagger$
($\tilde\op_1\tilde\op_1^\dagger$), reproduces the results of
Ref.\,\cite{Hankele:2006ja}.  In the scalar as well as in the
pseudo-scalar case the $\op_3$ ($\tilde\op_3$) induced term has a much
stronger variation than the formally leading $\op_1\op_1^\dagger$ and
the interference terms $\op_1\op_2^\dagger$ and $\op_1\op_5^\dagger$
(respectively the corresponding pseudo-scalar terms).  In the scalar
case, the ``4-quark-operator'' $\op_4$ leads to a remarkable deviation
from the other terms. As before, ``sum'' refers to the sum over the
partonic sub-channels for fixed $ij$.

Using {\tt vbfnlo}\,\cite{Arnold:2012xn,Arnold:2011wj,Arnold:2008rz}, we
calculated the $\Delta\Phi_{jj}$ distributions for top-loop induced Higgs
production. We find that they hardly differ from the results for
$\op_1^\dagger\op_1$ (respectively $\tilde\op_1^\dagger\tilde\op_1$)
and therefore refrain from including them in our plots.

An explanation for the different curvatures of the leading scalar and
pseudo-scalar operators $\op_1$ and $\tilde\op_1$, i.e. the suppression
of planar events for CP-odd couplings, is that the epsilon tensor
contracted with (four) linearly dependent momentum vectors of the
incoming and outgoing partons vanishes \cite{Hankele:2006ja}.

In order to allow the reader to derive quantitative results for
particular models corresponding to specific values for the Wilson
coefficients, we provide again in \fref{tab:phiij} and
\fref{tab:phiijtilde} the integrated cross sections derived with the
inclusive plus {\abbrev WBF} cuts described in \eqns{eq:inccuts}
and~(\ref{eq:wbfcuts}).

\begin{figure}
\includegraphics[width=\textwidth]{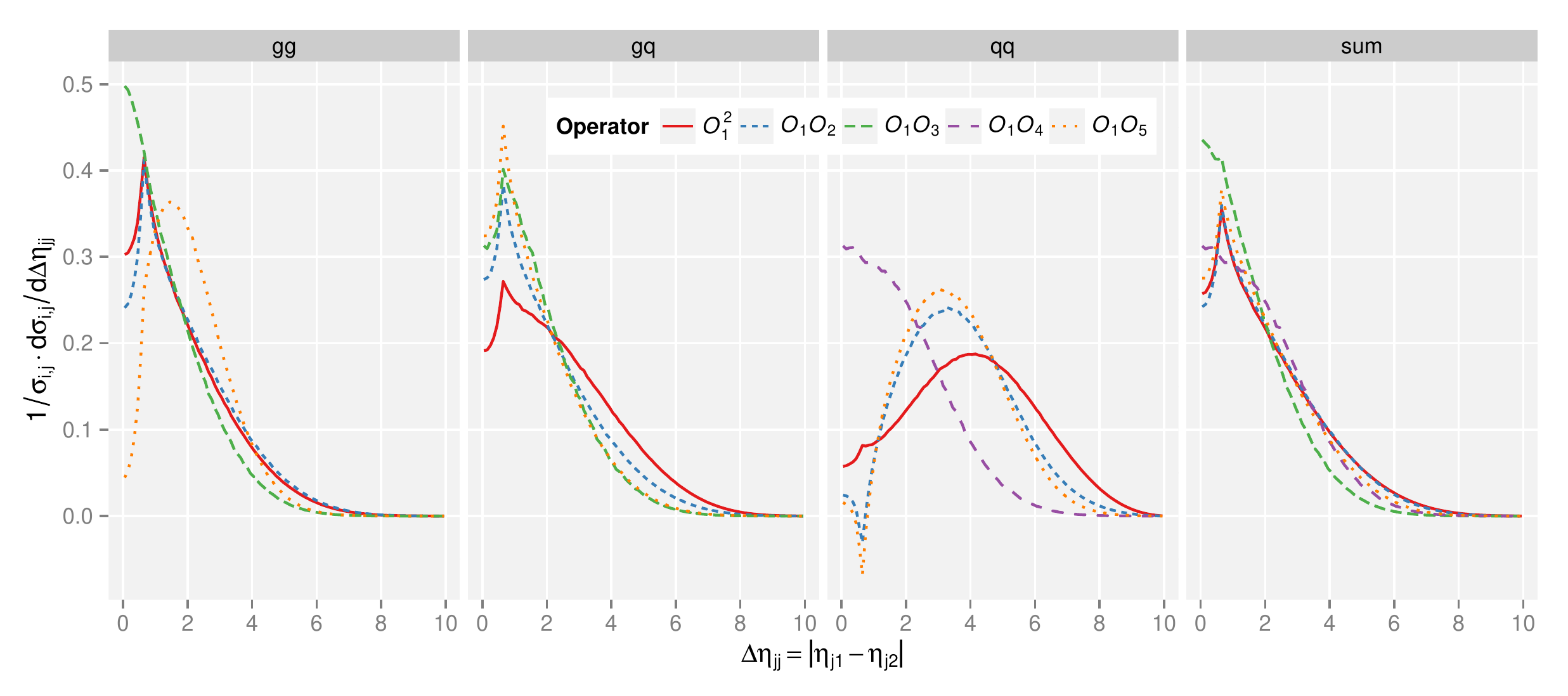}
\caption{Normalized distributions for rapidity separation of the two
  final state jets for scalar operators. The $\sigma_{ij}$ are given in
  \fref{tab:etajj}.}
\label{fig:etadist-LO-suppression}
\end{figure}

\begin{figure}
\includegraphics[width=\textwidth]{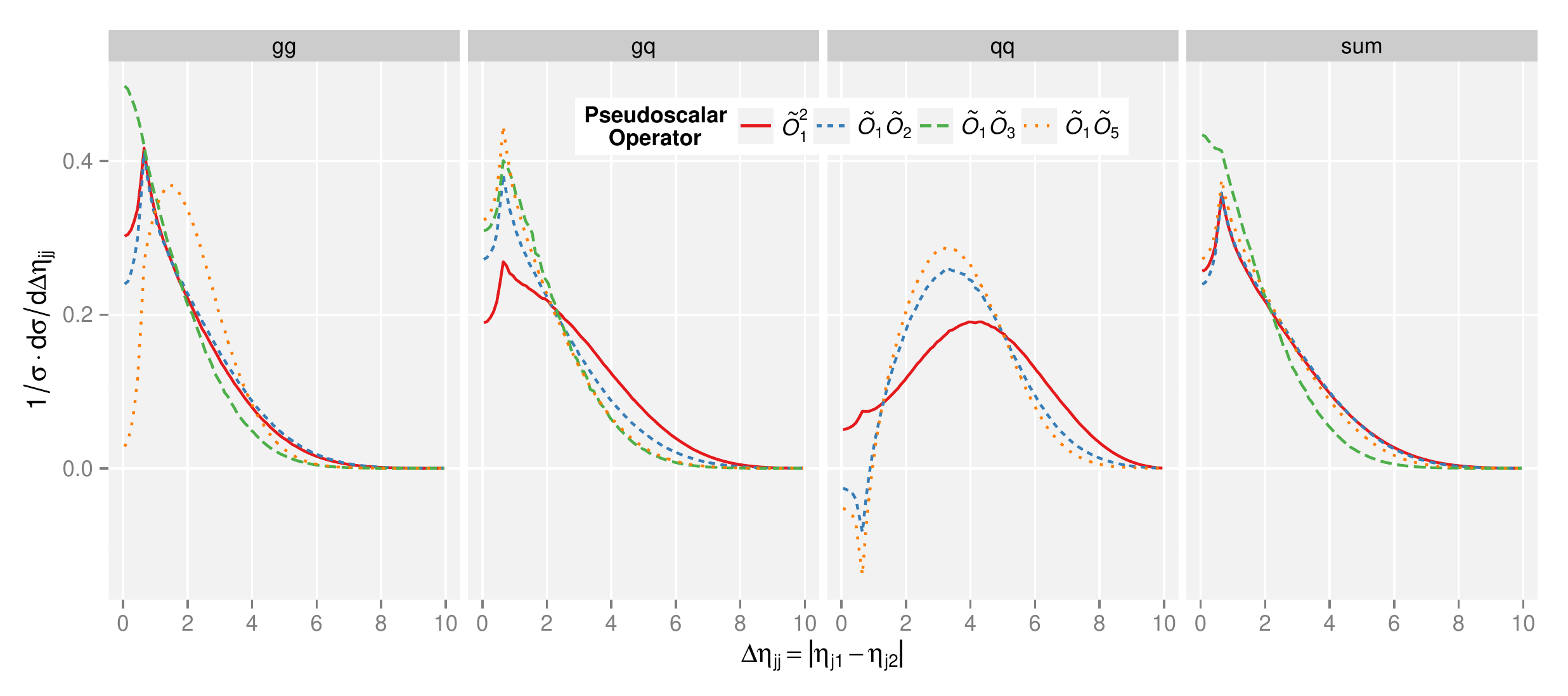}
\caption{Same as \fref{fig:etadist-LO-suppression}, but for
  pseudo-scalar operators. The $\sigma_{ij}$ are given in
  \fref{tab:etajjtilde}.}
\label{fig:etadist-pseudo-LO-suppression}
\end{figure}

Distributions for the jet rapidity separation for scalar and
pseudo-scalar operators are shown in \fref{fig:etadist-LO-suppression}
and~\ref{fig:etadist-pseudo-LO-suppression}. The used normalization
factors are given in \fref{tab:etajj} and \fref{tab:etajjtilde}. One can
compare these with the results of \cite[Figure 8]{DelDuca:2001fn}, where
the specifics for the $\Delta\eta_{jj}$ distribution for 2-jet
gluon-fusion and 2-jet weak boson fusion are discussed: while
gluon-fusion exhibits a peak at small $\Delta\eta_{jj}$ (due to the jet
radius constraint $R_{jj}>0.6$), for weak boson fusion the peak is at a
rapidity separation $\Delta\eta_{jj}\approx 5$ and considerably smaller.
We compared our results again to top-loop induced Higgs production
obtained with {\tt
  vbfnlo}\,\cite{Arnold:2012xn,Arnold:2011wj,Arnold:2008rz}; similar to
the $\Delta\Phi_{jj}$ distribution, we find that they are almost
identical to the curves for $\op_1^\dagger\op_1$ (respectively
$\tilde\op_1^\dagger\tilde\op_1$), which is why we refrain from
including them in our plots.

While there are quantitative differences among the various contributions
for the scalar and pseudo-scalar operators considered here, we conclude
that the qualitative differences are probably too small to be used in an
experimental analysis in order to classify the gluon-Higgs coupling.

\begin{figure}
\includegraphics[width=\textwidth]{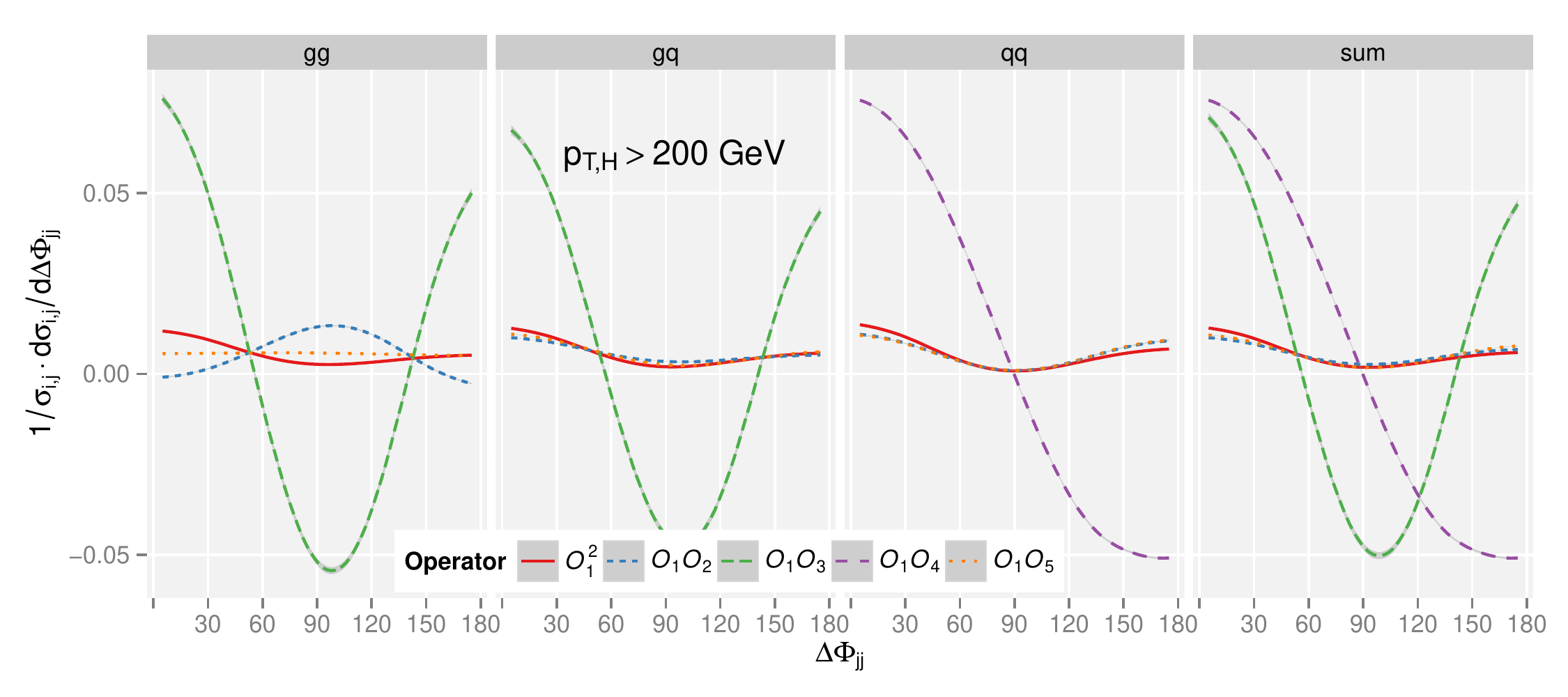}
\caption{Similar to \fref{fig:phidist-LO-suppression}, but restricted to
  events with $\pt>200$\,GeV, where $\pt$ is the transverse momentum of
  the Higgs boson.  The $\sigma_{ij}$ are given in
  \fref{tab:phiijpt200}.}
\label{fig:phidist-LO-suppression-pt200}
\end{figure}

\begin{figure}
\includegraphics[width=\textwidth]{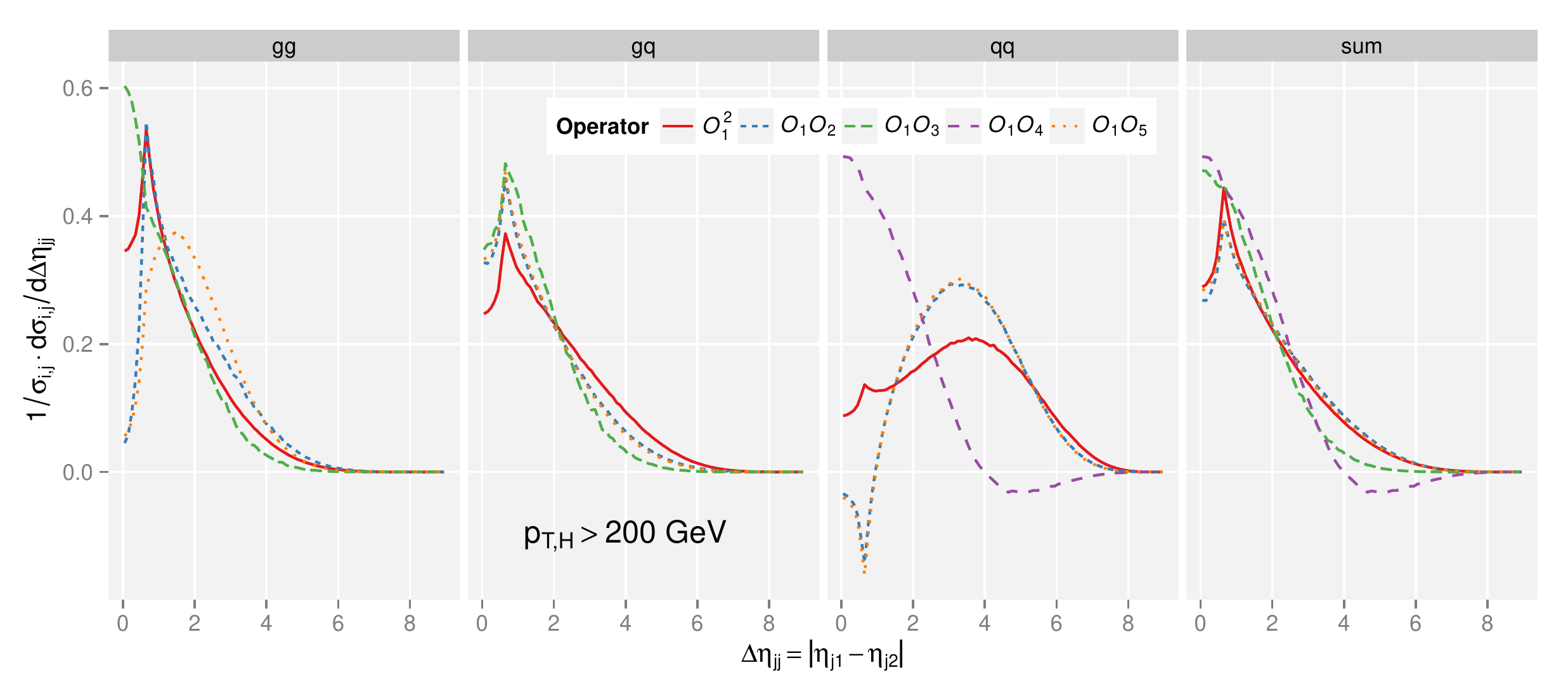}
\caption{Similar to \fref{fig:etadist-LO-suppression}, but restricted to
  events with $\pt>200$\,GeV, where $\pt$ is the transverse momentum of
  the Higgs boson.  The $\sigma_{ij}$ are given in
  \fref{tab:etaijpt200}.}
\label{fig:etadist-LO-suppression-pt200}
\end{figure}

In light of the fact that the differences among the various operators
increase with the Higgs transverse momentum, see
\fref{fig:ptdist-LO-suppression}
and~\ref{fig:ptdist_pseudo-LO-suppression}, it is suggestive to consider
the $\Delta\Phi_{jj}$ and $\Delta\eta_{jj}$ distributions for these
high-$\pt$ events only. For example,
\fref{fig:phidist-LO-suppression-pt200} shows the result for the
$\Delta\Phi_{jj}$ distribution in the scalar Higgs case, when the Higgs'
transverse momentum is restricted to $\pt>200$\,GeV.  Compared to
\fref{fig:phidist-LO-suppression}, some of the features are strongly
enhanced;\footnote{The fact that the slope of ``$\op_1\op_4$'' changes
  sign is due to a change of sign in the normalization, see
  \fref{tab:phiijpt200}.} however, considering the fact that
such a cut will signficantly decrease the data sample, it remains to be
seen whether it would lead to an improvement of an experimental
analysis.



\section{Higher order suppressed terms}\label{sec:higher}

Up to this point, we considered the Lagrangians in \eqn{eq:leff} and
\eqn{eq:lefftilde} as an effective theory truncated at
$\order{1/\Lambda^3}$. Therefore, we only took into account the square
of $\op_1$ (or $\tilde\op_1$) and its interference with $\op_n$ (or
$\tilde\op_n$) ($n\geq 2$), as other terms are of higher order in
$1/\Lambda$. In this section, we will consider products of the $\op_n$
(and $\tilde\op_n$) ($n\geq 2$) with each other. They might be relevant
if indeed the gluon-Higgs coupling is predominantly mediated by one of
the dimension-7 operators, as hypothesized in the introduction.  This
hypothesis is also the reason why we do not include interference terms
of $\op_1$ with dimension-9 operators in this section, as it would be
required if we were dealing with a regular analysis at fixed order in
$1/\Lambda$.

\begin{figure}
\includegraphics[width=\textwidth]{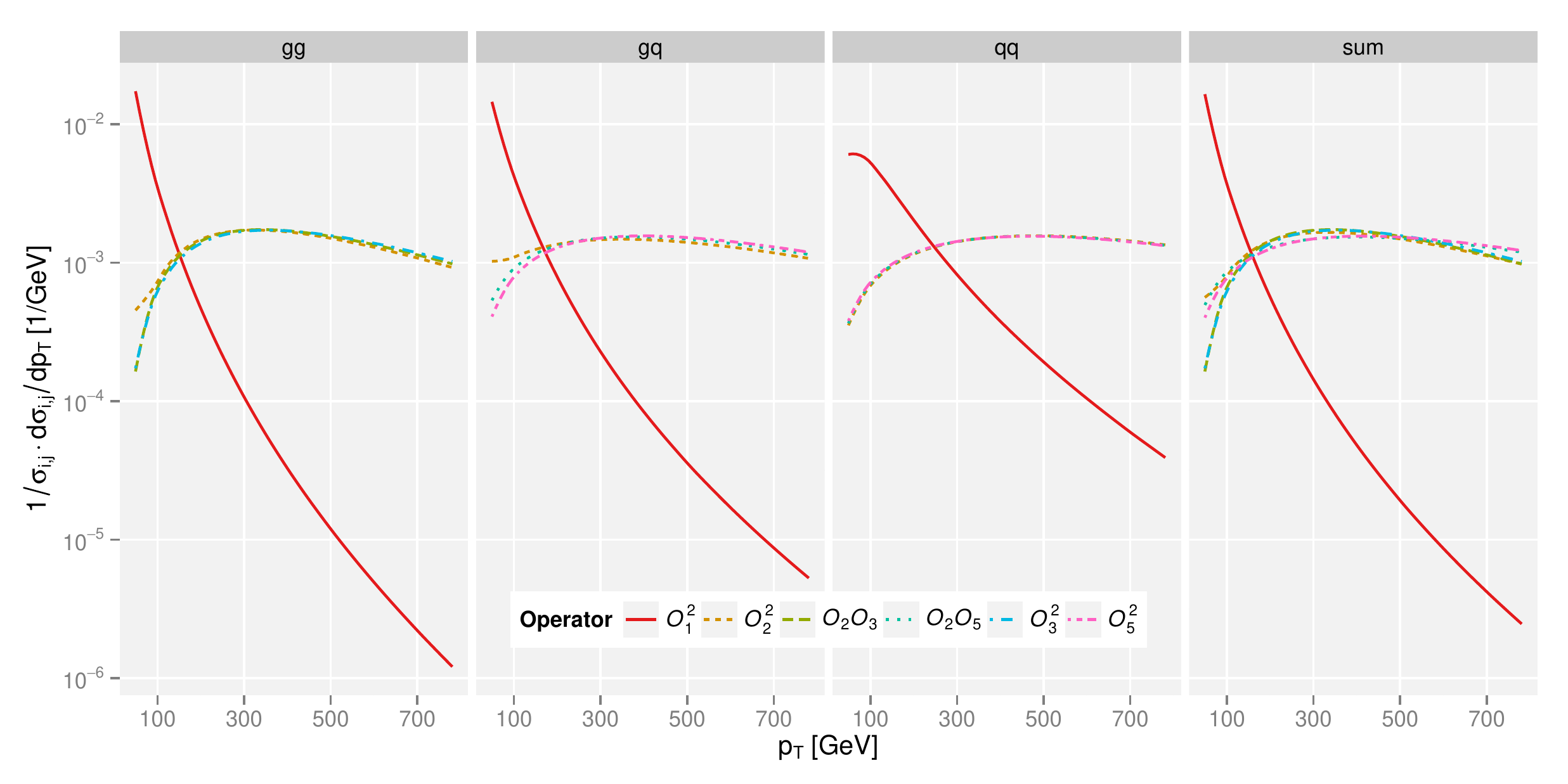}
\caption{Normalized Higgs transverse momentum distributions for scalar
  coupling operators. The $\sigma_{ij}$ are given in
  \fref{tab:sigmaijpthigh}.}
\label{fig:ptdist-high-high}
\end{figure}

\begin{figure}
\includegraphics[width=\textwidth]{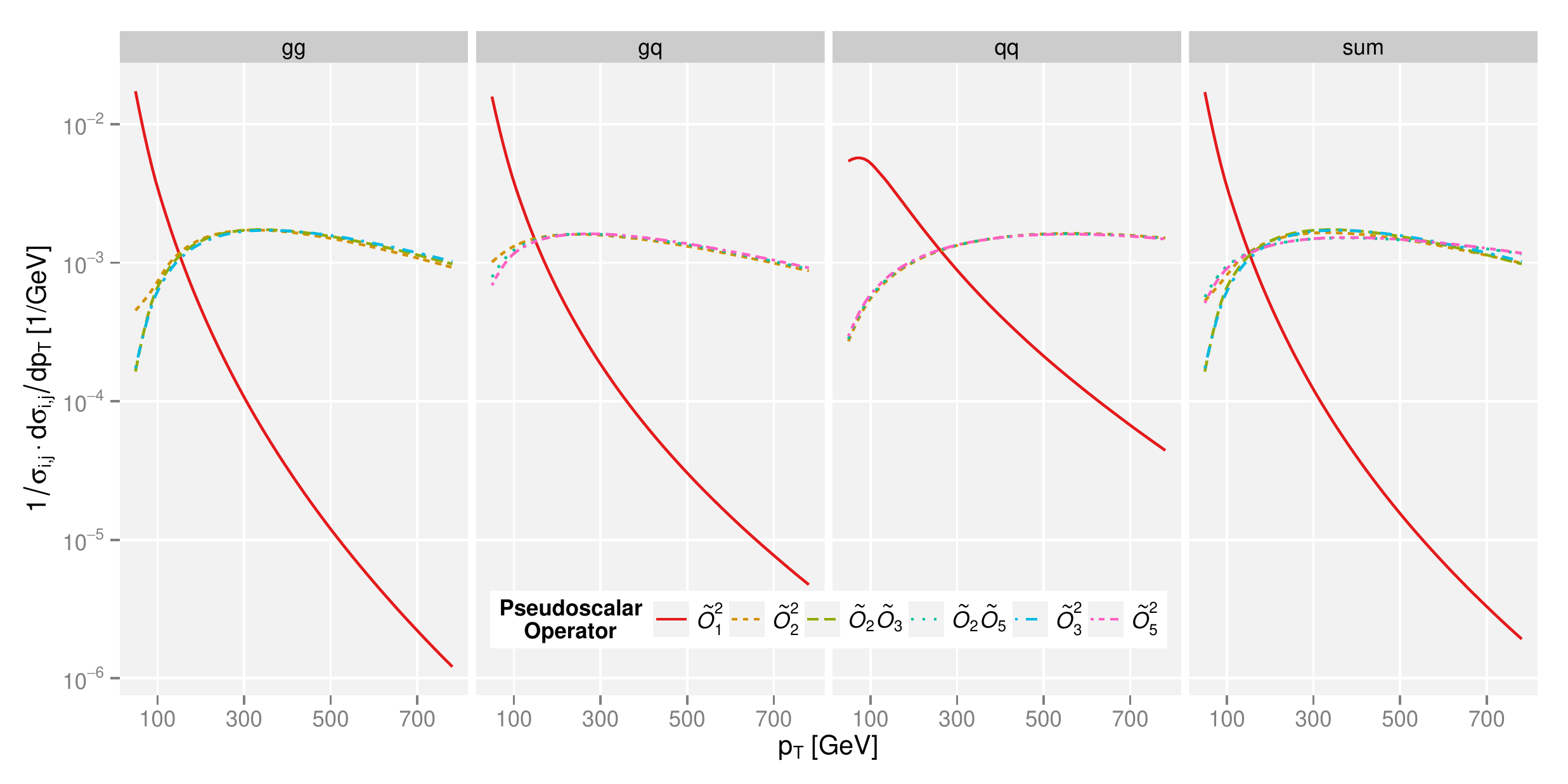}
\caption{Normalized Higgs transverse momentum distributions for
  pseudo-scalar coupling operators.  Note that the $gg$ channel is
  identical to the scalar case. The $\sigma_{ij}$ are given in
  \fref{tab:sigmaijpttildehigh}.
}
\label{fig:ptdist_pseudo-high-high}
\end{figure}

\Fref{fig:ptdist-high-high} and \ref{fig:ptdist_pseudo-high-high} show
the $\pt{}$-distributions in the $H$+1-jet case as induced by the higher
order operators. For comparison, we also include the formally leading term
arising from $\op_1\op_1^\dagger$. Two observations in these figures are
remarkable: on the one hand, the difference between the shapes of the
higher order terms and this formally leading term is quite
significant. On the other hand, the higher order terms themselves are
all very close to each other. Also, the behavior is very similar in the
scalar and the pseudo-scalar case.

The normalizations for the individual curves are given in
\fref{tab:sigmaijpthigh} and \ref{tab:sigmaijpttildehigh}.

\begin{figure}
\includegraphics[width=\textwidth]{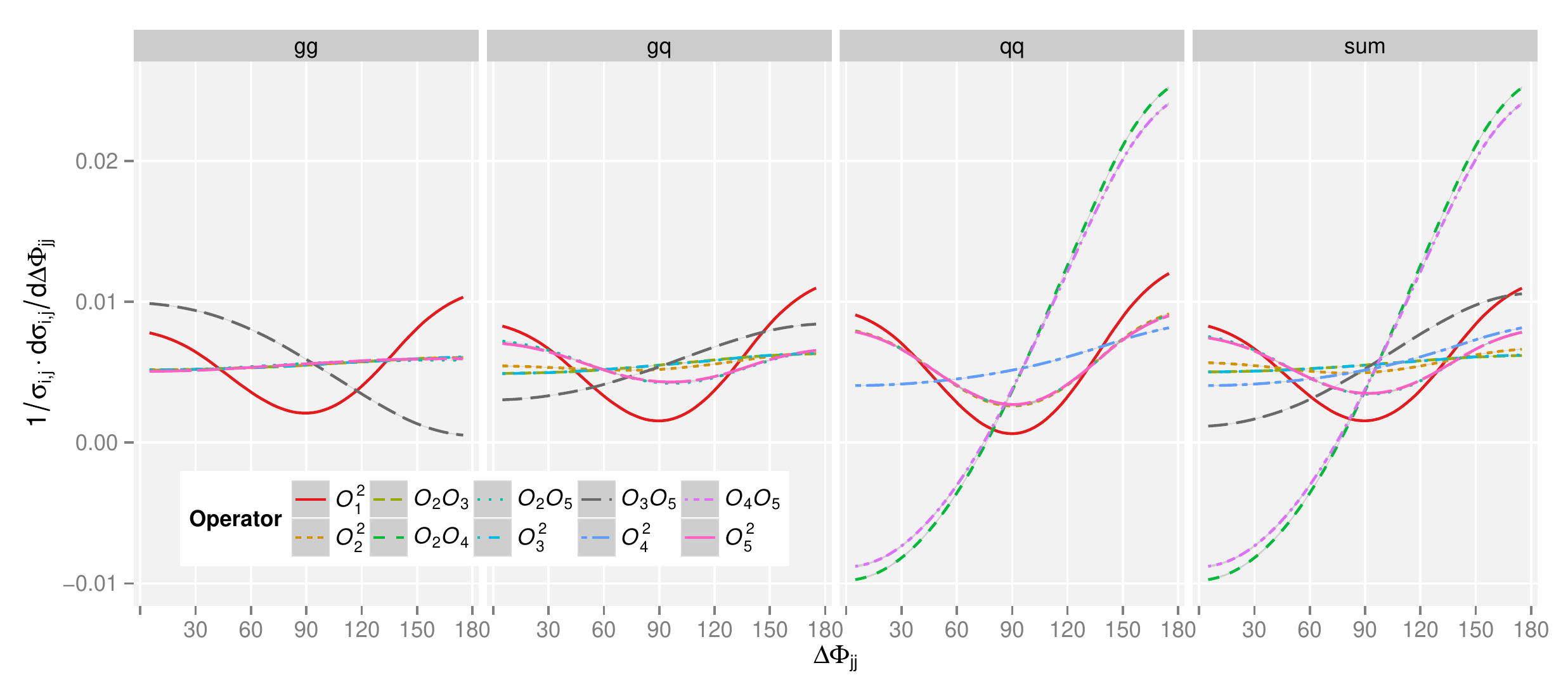}
\caption{\label{fig:phidist-LO-high}
Normalized distributions for azimuthal angle difference of the
  two final state jets for scalar operators. The $\sigma_{ij}$ are given in
  \fref{tab:phiijhigh}.
 }
\end{figure}

\begin{figure}
\includegraphics[width=\textwidth]{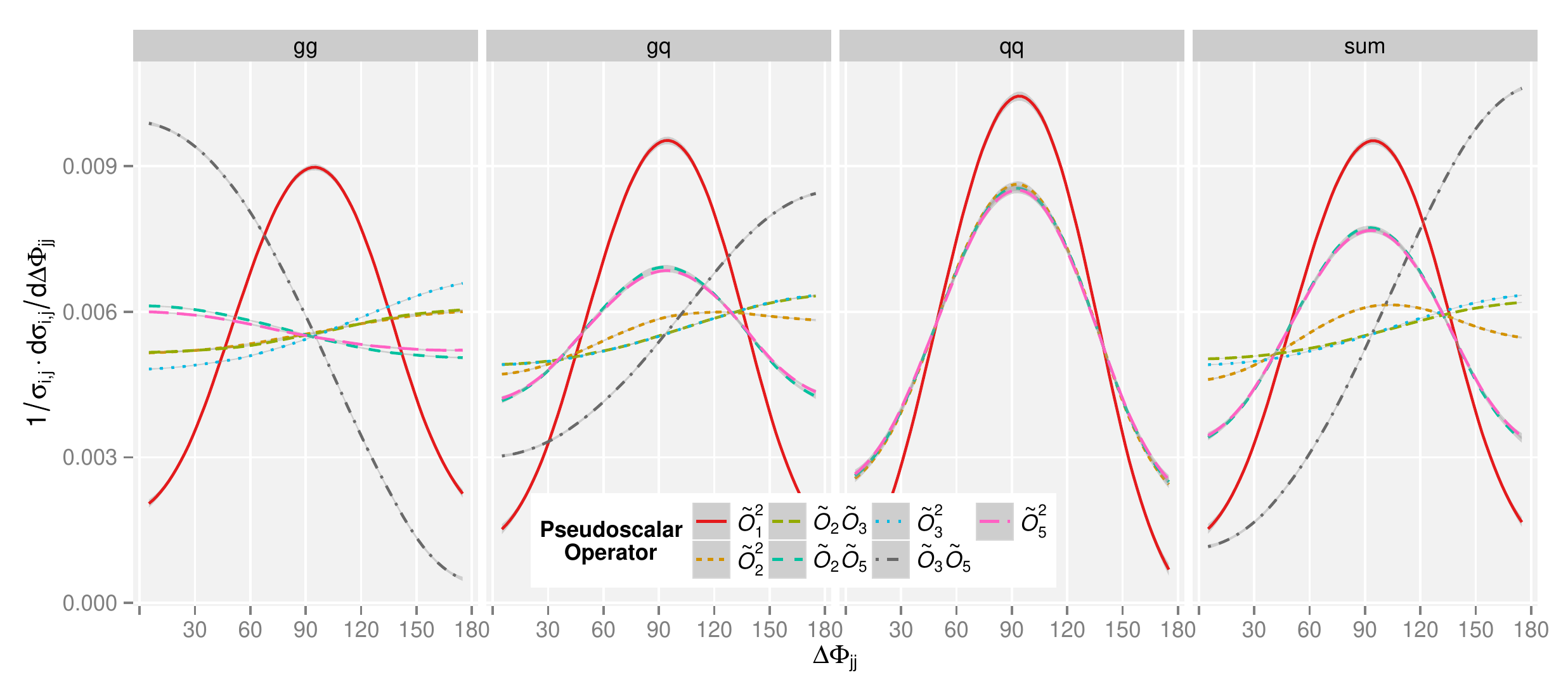}
\caption{Normalized distributions for azimuthal angle difference of the
  two final state jets for pseudo-scalar operators.  The $\sigma_{ij}$ are given in  \fref{tab:phiijtildehigh}.
}
\label{fig:phidist-pseudo-high-high}
\end{figure}

\begin{figure}[h]
\includegraphics[width=\textwidth]{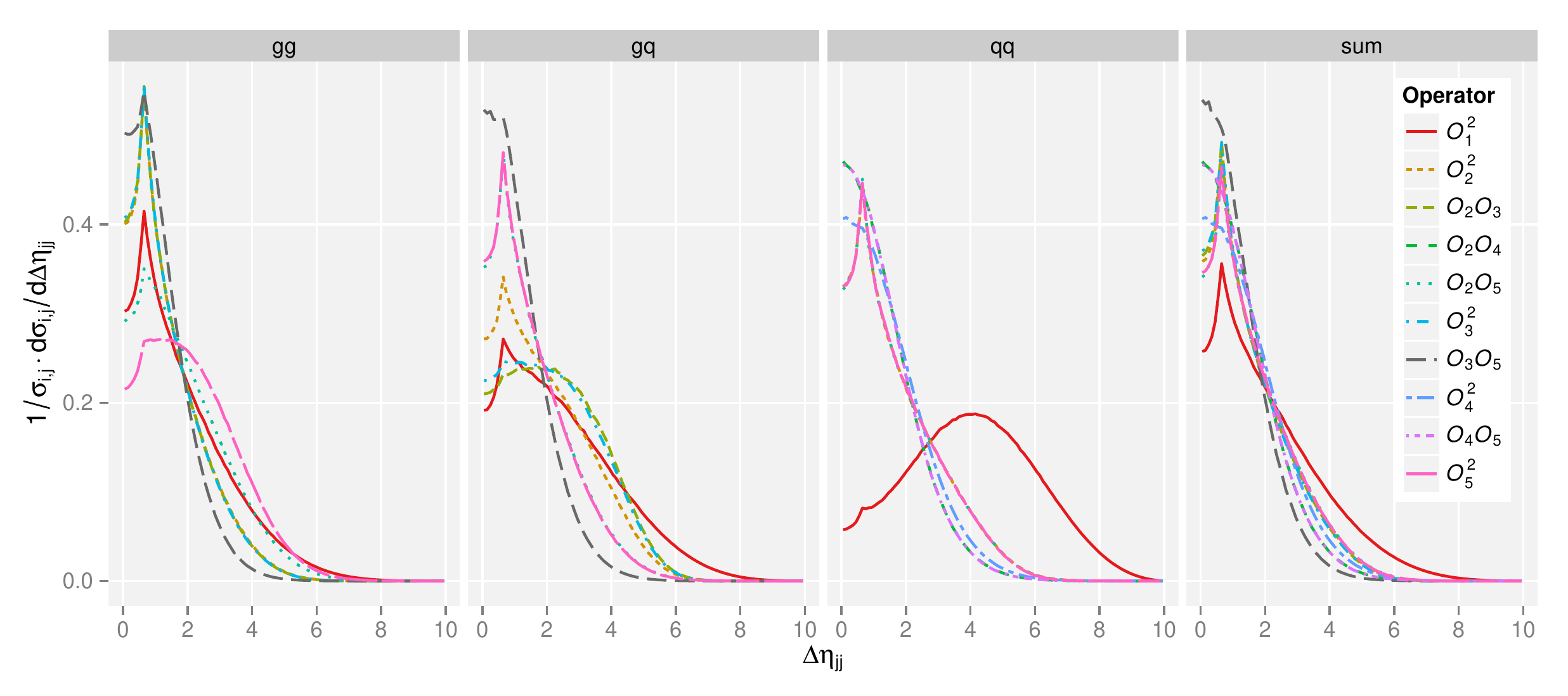}
\caption{Normalized distributions for rapidity separation of the two
  final state jets for scalar operators.  The $\sigma_{ij}$ are given in
  \fref{tab:etajjhigh}.
}
\label{fig:etadist-high-high}
\end{figure}

\begin{figure}[h]
\includegraphics[width=\textwidth]{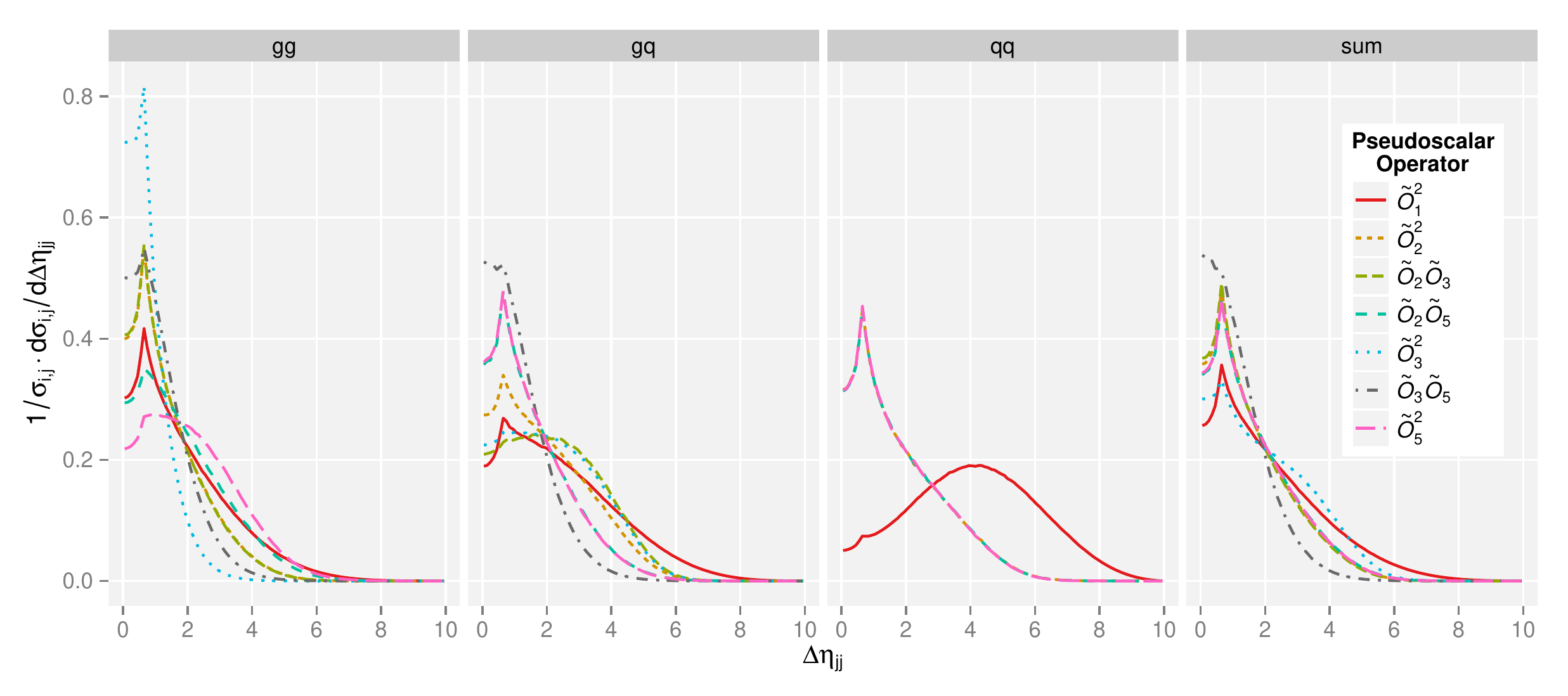}
\caption{Normalized distributions for rapidity separation of the two
  final state jets for pseudo-scalar operators.  The $\sigma_{ij}$ are given in
  \fref{tab:etajjtildehigh}.
}
\label{fig:etadist-pseudo-high-high}
\end{figure}

Similarly, we consider the $\Delta\Phi_{jj}$ distributions for the
higher dimensional operators in \fref{fig:phidist-LO-high} and
\ref{fig:phidist-pseudo-high-high}, and the $\Delta\eta_{jj}$
distributions in \fref{fig:etadist-high-high} and
\ref{fig:etadist-pseudo-high-high}. While there are quite large
qualitative differences in the various $\Delta\Phi_{jj}$ distributions,
they are far less prominent in the $\Delta\eta_{jj}$ shapes. The
formally leading term induced by $\op_1\op_1^\dagger$ (and
$\tilde\op_1\tilde\op_1^\dagger$) are again included for comparison.

The corresponding normalization factors are given in
\fref{tab:phiijhigh}, \ref{tab:phiijtildehigh} and \fref{tab:etajjhigh}, 
\ref{tab:etajjtildehigh}.

\FloatBarrier

\section{Conclusions}\label{sec:conclusions}

In this paper we have studied the influence of higher dimensional
operators that couple gluons to a scalar or a pseudo-scalar particle $H$
on distributions in $H$+1-jet and $H$+2-jets production. While the \sm{}
assumes a (mostly) top-loop induced gluon-Higgs coupling, an analysis
along the lines of our paper will be necessary in order to fully test
this hypothesis. We have found that dimension-7 operators significantly
affect the $\pt{}$ distribution of the Higgs boson, leading to a much
harder spectrum than in the \sm{}. Also 2-jet observables can be quite
sensitive to these effects that are formally subleading in the scale of
``new physics''.

We note that since, to our knowledge, this is the first analysis of this
kind, we have restricted ourselves to the lowest order of perturbation
theory. If deviations in the observables discussed here relative to the
\sm{} prediction are observed, it will be necessary and interesting to
perform this analysis at \nlo{} \qcd{}.


\paragraph{Acknowledgments}

We would like to thank D.\,Zeppenfeld for comments on the manuscript and
calling our attention to Ref.\,\cite{Germer:diplom} which considers the
effect of dimension-7 operators to \sm{} Higgs production. Further
thanks go to H.\,Mantler for valuable help with the \sm{}
predictions. This work was supported by {\abbrev BMBF}, contract
{\abbrev 05H12PXE}.


\begin{appendix}

\section{Normalization factors}\label{app:norm}
In this appendix, we collect the normalization factors for the
distributions calculated in this paper. They should allow the reader to
reconstruct the absolute distributions for any particular model with
specific values of the Wilson coefficients $C_i$ or $\tilde C_i$, see
\eqns{eq:leff} and (\ref{eq:lefftilde}). Specifically, absolute cross
sections $\dd\sigma_{ij}$ in picobarns are obtained by multiplying the
numbers for $\dd\sigma_{ij}/\sigma_{ij}$ read off from the figures in
Sections~\ref{sec:distros} and \ref{sec:higher} by the normalization
factors $\sigma_{ij}$ given in the tables below, times
$\text{Re}(C_i^\dagger C_j)/(\Lambda/\text{GeV})^{(n_i+n_j)}$, where
$n_1=1$ and $n_i=3$ for $i\neq 1$.\footnote{This factor is not required
  for the cross sections labelled ``\sm'' or ``top-loop'', of course.}
In other words, the numbers given here for the $\sigma_{ij}$ refer to
$\Lambda=1$\,GeV and $C_i=1\ \forall i$.


\begin{table}
  \begin{center}
\begin{tabular}{|c|c|c|c|}\hline \multicolumn{4}{|c|}{$\sigma{}_{ij}$/pb for $H$+1-jet (scalar)}\tabularnewline\hline $ij$&$gg$&$gq$&$qq$\tabularnewline\hline \hline \sm{}$/\alpha{}_{s}^{2}$&$4.79 \cdot 10^{2}$&$1.78 \cdot 10^{2}$&$3.13$\tabularnewline\hline 11&$3.88 \cdot 10^{10}$&$1.59 \cdot 10^{10}$&$1.03 \cdot 10^{8}$\tabularnewline\hline 12&$-5.55 \cdot 10^{14}$&$-6.59 \cdot 10^{14}$&$3.05 \cdot 10^{13}$\tabularnewline\hline 13&$-2.00 \cdot 10^{13}$&--&--\tabularnewline\hline 15&--&$-2.05 \cdot 10^{14}$&$1.60 \cdot 10^{13}$\tabularnewline\hline \end{tabular}
    \caption[]{\label{tab:sigmaijpt} Normalization factors for $\pt{}$
      distributions in $H$+1-jet production for a scalar Higgs. They are
      obtained by integrating the distributions of
      \fref{fig:ptdist-LO-suppression} over the interval
      $\pt\in[30,800]$\,GeV.}
  \end{center}
\end{table}


\begin{table}
\begin{center}
\begin{tabular}{|c|c|c|c|}\hline \multicolumn{4}{|c|}{$\sigma{}_{ij}$/pb for $H$+1-jet (pseudo-scalar)}\tabularnewline\hline $ij$&$gg$&$gq$&$qq$\tabularnewline\hline \hline top-loop$/\alpha{}_{s}^{2}$&$1.11 \cdot 10^{3}$&$4.16 \cdot 10^{2}$&$6.67$\tabularnewline\hline 11&$3.88 \cdot 10^{10}$&$5.06 \cdot 10^{9}$&$3.13 \cdot 10^{8}$\tabularnewline\hline 12&$-5.56 \cdot 10^{14}$&$-3.22 \cdot 10^{14}$&$8.87 \cdot 10^{13}$\tabularnewline\hline 13&$-2.00 \cdot 10^{13}$&--&--\tabularnewline\hline 15&--&$-1.21 \cdot 10^{14}$&$4.68 \cdot 10^{13}$\tabularnewline\hline \end{tabular}
\caption[]{\label{tab:sigmaijpttilde}Same as table\,\ref{tab:sigmaijpt}, but
  for a pseudo-scalar Higgs (see \fref{fig:ptdist_pseudo-LO-suppression}).}
\end{center}
\end{table}


\begin{table}
\begin{center}
\begin{tabular}{|c|c|c|c|}\hline \multicolumn{4}{|c|}{$\sigma{}_{ij}$/pb for $H$+2-jet (scalar), WBF cuts}\tabularnewline\hline $ij$&$gg$&$gq$&$qq$\tabularnewline\hline \hline 11&$1.43 \cdot 10^{9}$&$2.27 \cdot 10^{9}$&$8.42 \cdot 10^{8}$\tabularnewline\hline 12&$-2.14 \cdot 10^{13}$&$-6.54 \cdot 10^{13}$&$-4.47 \cdot 10^{13}$\tabularnewline\hline 13&$-7.13 \cdot 10^{11}$&$-7.59 \cdot 10^{11}$&--\tabularnewline\hline 14&--&--&$-1.19 \cdot 10^{12}$\tabularnewline\hline 15&$-4.65 \cdot 10^{11}$&$-1.60 \cdot 10^{13}$&$-1.58 \cdot 10^{13}$\tabularnewline\hline \end{tabular}
\caption[]{\label{tab:phiij}Normalization factors for $\Delta\Phi_{jj}$
  distributions in $H$+2-jets production for a scalar Higgs. They are
  obtained by integrating the distributions of
  \fref{fig:phidist-LO-suppression} over the interval
  $\Delta\Phi_{jj}\in[0,\pi]$, with the cuts described in
  \eqns{eq:inccuts} and (\ref{eq:wbfcuts}).}
\end{center}
\end{table}


\begin{table}
\begin{center}
\begin{tabular}{|c|c|c|c|}\hline \multicolumn{4}{|c|}{$\sigma{}_{ij}$/pb for $H$+2-jet (pseudo-scalar), WBF cuts}\tabularnewline\hline $ij$&$gg$&$gq$&$qq$\tabularnewline\hline \hline 11&$1.42 \cdot 10^{9}$&$2.24 \cdot 10^{9}$&$8.29 \cdot 10^{8}$\tabularnewline\hline 12&$-2.11 \cdot 10^{13}$&$-6.66 \cdot 10^{13}$&$-4.41 \cdot 10^{13}$\tabularnewline\hline 13&$-7.12 \cdot 10^{11}$&$-7.57 \cdot 10^{11}$&--\tabularnewline\hline 14&--&--&--\tabularnewline\hline 15&$-4.43 \cdot 10^{11}$&$-1.68 \cdot 10^{13}$&$-1.55 \cdot 10^{13}$\tabularnewline\hline \end{tabular}
\end{center}
\caption[]{\label{tab:phiijtilde}Same as \fref{tab:phiij}, but for a
  pseudo-scalar Higgs (see \fref{fig:phidist-pseudo-LO-suppression}).}
\end{table}


\begin{table}
\begin{center}
\begin{tabular}{|c|c|c|c|}\hline \multicolumn{4}{|c|}{$\sigma{}_{ij}$/pb for $H$+2-jet (scalar), incl. cuts}\tabularnewline\hline $ij$&$gg$&$gq$&$qq$\tabularnewline\hline \hline 11&$5.00 \cdot 10^{10}$&$2.67 \cdot 10^{10}$&$2.50 \cdot 10^{9}$\tabularnewline\hline 12&$-6.41 \cdot 10^{14}$&$-1.05 \cdot 10^{15}$&$-1.48 \cdot 10^{14}$\tabularnewline\hline 13&$-6.11 \cdot 10^{13}$&$-3.11 \cdot 10^{13}$&--\tabularnewline\hline 14&--&--&$-1.71 \cdot 10^{13}$\tabularnewline\hline 15&$-1.42 \cdot 10^{13}$&$-3.59 \cdot 10^{14}$&$-5.51 \cdot 10^{13}$\tabularnewline\hline \end{tabular}
\caption[]{\label{tab:etajj}Normalization factors for $\Delta\eta_{jj}$
  distributions in $H$+2-jets production for a scalar Higgs. They are
  obtained by integrating the distributions of
  \fref{fig:etadist-LO-suppression} over the interval
  $\Delta\eta_{jj}\in[0,10]$, with the cuts described in
  \eqns{eq:inccuts}.}
\end{center}
\end{table}


\begin{table}
\begin{center}
\begin{tabular}{|c|c|c|c|}\hline \multicolumn{4}{|c|}{$\sigma{}_{ij}$/pb for $H$+2-jet (pseudo-scalar), incl. cuts}\tabularnewline\hline $ij$&$gg$&$gq$&$qq$\tabularnewline\hline \hline 11&$4.94 \cdot 10^{10}$&$2.62 \cdot 10^{10}$&$2.39 \cdot 10^{9}$\tabularnewline\hline 12&$-6.32 \cdot 10^{14}$&$-1.07 \cdot 10^{15}$&$-1.31 \cdot 10^{14}$\tabularnewline\hline 13&$-6.11 \cdot 10^{13}$&$-3.11 \cdot 10^{13}$&--\tabularnewline\hline 14&--&--&--\tabularnewline\hline 15&$-1.36 \cdot 10^{13}$&$-3.70 \cdot 10^{14}$&$-4.74 \cdot 10^{13}$\tabularnewline\hline \end{tabular}
\end{center}
\caption[]{\label{tab:etajjtilde}Same as \fref{tab:etajj}, but for a
  pseudo-scalar Higgs (see \fref{fig:etadist-pseudo-LO-suppression}).}
\end{table}


\begin{table}
\begin{center}
\begin{tabular}{|c|c|c|c|}\hline \multicolumn{4}{|c|}{\begin{tabular}[c]{@{}c@{}}$\sigma{}_{ij}$/pb for $H$+2-jet (scalar), WBF cuts\\ $p_{\mathrm{T,H}}>200\,\mathrm{GeV}$\end{tabular}}\tabularnewline\hline $ij$&$gg$&$gq$&$qq$\tabularnewline\hline \hline 11&$1.14 \cdot 10^{8}$&$2.36 \cdot 10^{8}$&$1.23 \cdot 10^{8}$\tabularnewline\hline 12&$-1.91 \cdot 10^{12}$&$-2.32 \cdot 10^{13}$&$-2.25 \cdot 10^{13}$\tabularnewline\hline 13&$-1.28 \cdot 10^{11}$&$-1.93 \cdot 10^{11}$&--\tabularnewline\hline 14&--&--&$3.82 \cdot 10^{11}$\tabularnewline\hline 15&$-2.34 \cdot 10^{11}$&$-1.00 \cdot 10^{13}$&$-1.03 \cdot 10^{13}$\tabularnewline\hline \end{tabular}
\caption[]{\label{tab:phiijpt200}Normalization factors for
  $\Delta\Phi_{jj}$ distributions in $H$+2-jets production for a scalar
  Higgs with $\pt>200$\,GeV. They are obtained by integrating the
  distributions of \fref{fig:phidist-LO-suppression-pt200} over the
  interval $\Delta\Phi_{jj}\in[0,\pi]$, with the cuts described in
  \eqns{eq:inccuts} and (\ref{eq:wbfcuts}).}
\end{center}
\end{table}


\begin{table}
\begin{center}
\begin{tabular}{|c|c|c|c|}\hline \multicolumn{4}{|c|}{\begin{tabular}[c]{@{}c@{}}$\sigma{}_{ij}$/pb for $H$+2-jet (scalar), incl. cuts\\ $p_{\mathrm{T,H}}>200\,\mathrm{GeV}$\end{tabular}}\tabularnewline\hline $ij$&$gg$&$gq$&$qq$\tabularnewline\hline \hline 11&$3.58 \cdot 10^{9}$&$2.95 \cdot 10^{9}$&$3.84 \cdot 10^{8}$\tabularnewline\hline 12&$-3.83 \cdot 10^{13}$&$-5.05 \cdot 10^{14}$&$-7.00 \cdot 10^{13}$\tabularnewline\hline 13&$-1.21 \cdot 10^{13}$&$-1.30 \cdot 10^{13}$&--\tabularnewline\hline 14&--&--&$-6.14 \cdot 10^{12}$\tabularnewline\hline 15&$-6.57 \cdot 10^{12}$&$-2.46 \cdot 10^{14}$&$-3.23 \cdot 10^{13}$\tabularnewline\hline \end{tabular}
\caption[]{\label{tab:etaijpt200}Normalization factors for
  $\Delta\eta_{jj}$ distributions in $H$+2-jets production for a scalar
  Higgs with $\pt>200$\,GeV. They are
  obtained by integrating the distributions of
  \fref{fig:etadist-LO-suppression} over the interval
  $\Delta\eta_{jj}\in[0,10]$, with the cuts described in
  \eqns{eq:inccuts}.}
\end{center}
\end{table}


\begin{table}
  \begin{center}
\begin{tabular}{|c|c|c|c|}\hline \multicolumn{4}{|c|}{$\sigma{}_{ij}$/pb for $H$+1-jet (scalar)}\tabularnewline\hline $ij$&$gg$&$gq$&$qq$\tabularnewline\hline \hline 22&$1.35 \cdot 10^{20}$&$3.60 \cdot 10^{19}$&$1.00 \cdot 10^{19}$\tabularnewline\hline 23&$-1.12 \cdot 10^{20}$&--&--\tabularnewline\hline 25&--&$3.35 \cdot 10^{19}$&$1.01 \cdot 10^{19}$\tabularnewline\hline 33&$2.38 \cdot 10^{19}$&--&--\tabularnewline\hline 55&--&$7.97 \cdot 10^{18}$&$2.57 \cdot 10^{18}$\tabularnewline\hline \end{tabular}
    \caption[]{\label{tab:sigmaijpthigh} Normalization factors for
      $\pt{}$ distributions in $H$+1-jet production induced by higher
      dimensional operators for a scalar Higgs (see
      \fref{fig:ptdist-high-high}).} 
  \end{center}
\end{table}


\begin{table}
\begin{center}
\begin{tabular}{|c|c|c|c|}\hline \multicolumn{4}{|c|}{$\sigma{}_{ij}$/pb for $H$+1-jet (pseudo-scalar)}\tabularnewline\hline $ij$&$gg$&$gq$&$qq$\tabularnewline\hline \hline 22&$1.35 \cdot 10^{20}$&$3.61 \cdot 10^{19}$&$2.67 \cdot 10^{19}$\tabularnewline\hline 23&$-1.12 \cdot 10^{20}$&--&--\tabularnewline\hline 25&--&$3.48 \cdot 10^{19}$&$2.71 \cdot 10^{19}$\tabularnewline\hline 33&$2.38 \cdot 10^{19}$&--&--\tabularnewline\hline 55&--&$8.47 \cdot 10^{18}$&$6.87 \cdot 10^{18}$\tabularnewline\hline \end{tabular}
\caption[]{\label{tab:sigmaijpttildehigh}Same as
  table\,\ref{tab:sigmaijpthigh}, but for a pseudo-scalar Higgs (see
  \fref{fig:ptdist_pseudo-high-high}).}
\end{center}
\end{table}


\begin{table}
\begin{center}
\begin{tabular}{|c|c|c|c|}\hline \multicolumn{4}{|c|}{$\sigma{}_{ij}$/pb for $H$+2-jet (scalar), WBF cuts}\tabularnewline\hline $ij$&$gg$&$gq$&$qq$\tabularnewline\hline \hline 22&$1.40 \cdot 10^{19}$&$2.01 \cdot 10^{19}$&$5.14 \cdot 10^{18}$\tabularnewline\hline 23&$-1.07 \cdot 10^{19}$&$-1.24 \cdot 10^{19}$&--\tabularnewline\hline 24&--&--&$3.09 \cdot 10^{17}$\tabularnewline\hline 25&$2.28 \cdot 10^{17}$&$4.17 \cdot 10^{18}$&$4.96 \cdot 10^{18}$\tabularnewline\hline 33&$2.08 \cdot 10^{18}$&$2.45 \cdot 10^{18}$&--\tabularnewline\hline 34&--&--&--\tabularnewline\hline 35&$-2.06 \cdot 10^{15}$&$9.60 \cdot 10^{15}$&--\tabularnewline\hline 44&--&--&$1.20 \cdot 10^{17}$\tabularnewline\hline 45&--&--&$1.49 \cdot 10^{17}$\tabularnewline\hline 55&$5.48 \cdot 10^{16}$&$1.02 \cdot 10^{18}$&$1.21 \cdot 10^{18}$\tabularnewline\hline \end{tabular}
\caption[]{\label{tab:phiijhigh}Normalization factors for $\Delta\Phi_{jj}$
  distributions in $H$+2-jets production induced by higher dimensional
  operators for a scalar Higgs (see \fref{fig:phidist-LO-high}).}
\end{center}
\end{table}


\begin{table}
\begin{center}
\begin{tabular}{|c|c|c|c|}\hline \multicolumn{4}{|c|}{$\sigma{}_{ij}$/pb for $H$+2-jet (pseudo-scalar), WBF cuts}\tabularnewline\hline $ij$&$gg$&$gq$&$qq$\tabularnewline\hline \hline 22&$1.40 \cdot 10^{19}$&$2.01 \cdot 10^{19}$&$4.88 \cdot 10^{18}$\tabularnewline\hline 23&$-1.07 \cdot 10^{19}$&$-1.24 \cdot 10^{19}$&--\tabularnewline\hline 24&--&--&--\tabularnewline\hline 25&$2.28 \cdot 10^{17}$&$4.12 \cdot 10^{18}$&$4.71 \cdot 10^{18}$\tabularnewline\hline 33&$2.26 \cdot 10^{15}$&$2.45 \cdot 10^{18}$&--\tabularnewline\hline 34&--&--&--\tabularnewline\hline 35&$-2.06 \cdot 10^{15}$&$9.62 \cdot 10^{15}$&--\tabularnewline\hline 44&--&--&--\tabularnewline\hline 45&--&--&--\tabularnewline\hline 55&$5.49 \cdot 10^{16}$&$1.01 \cdot 10^{18}$&$1.15 \cdot 10^{18}$\tabularnewline\hline \end{tabular}
\end{center}
\caption[]{\label{tab:phiijtildehigh}Same as \fref{tab:phiijhigh}, but for a
  pseudo-scalar Higgs (see \fref{fig:phidist-pseudo-high-high}).}
\end{table}


\begin{table}
\begin{center}
\begin{tabular}{|c|c|c|c|}\hline \multicolumn{4}{|c|}{$\sigma{}_{ij}$/pb for $H$+2-jet (scalar), incl. cuts}\tabularnewline\hline $ij$&$gg$&$gq$&$qq$\tabularnewline\hline \hline 22&$7.10 \cdot 10^{20}$&$3.10 \cdot 10^{20}$&$9.56 \cdot 10^{19}$\tabularnewline\hline 23&$-5.62 \cdot 10^{20}$&$-1.36 \cdot 10^{20}$&--\tabularnewline\hline 24&--&--&$1.69 \cdot 10^{19}$\tabularnewline\hline 25&$4.21 \cdot 10^{18}$&$1.38 \cdot 10^{20}$&$9.38 \cdot 10^{19}$\tabularnewline\hline 33&$1.12 \cdot 10^{20}$&$2.86 \cdot 10^{19}$&--\tabularnewline\hline 34&--&--&--\tabularnewline\hline 35&$-4.94 \cdot 10^{17}$&$1.65 \cdot 10^{18}$&--\tabularnewline\hline 44&--&--&$4.26 \cdot 10^{18}$\tabularnewline\hline 45&--&--&$8.37 \cdot 10^{18}$\tabularnewline\hline 55&$7.18 \cdot 10^{17}$&$3.49 \cdot 10^{19}$&$2.32 \cdot 10^{19}$\tabularnewline\hline \end{tabular}
\caption[]{\label{tab:etajjhigh}Normalization factors for $\Delta\eta_{jj}$
  distributions in $H$+2-jets production induced by higher dimensional
  operators for a scalar Higgs (see \fref{fig:etadist-high-high}).}
\end{center}
\end{table}


\begin{table}
\begin{center}
\begin{tabular}{|c|c|c|c|}\hline \multicolumn{4}{|c|}{$\sigma{}_{ij}$/pb for $H$+2-jet (pseudo-scalar), incl. cuts}\tabularnewline\hline $ij$&$gg$&$gq$&$qq$\tabularnewline\hline \hline 22&$7.11 \cdot 10^{20}$&$3.10 \cdot 10^{20}$&$8.33 \cdot 10^{19}$\tabularnewline\hline 23&$-5.62 \cdot 10^{20}$&$-1.37 \cdot 10^{20}$&--\tabularnewline\hline 24&--&--&--\tabularnewline\hline 25&$4.19 \cdot 10^{18}$&$1.39 \cdot 10^{20}$&$8.15 \cdot 10^{19}$\tabularnewline\hline 33&$5.12 \cdot 10^{18}$&$2.86 \cdot 10^{19}$&--\tabularnewline\hline 34&--&--&--\tabularnewline\hline 35&$-4.94 \cdot 10^{17}$&$1.65 \cdot 10^{18}$&--\tabularnewline\hline 44&--&--&--\tabularnewline\hline 45&--&--&--\tabularnewline\hline 55&$7.15 \cdot 10^{17}$&$3.50 \cdot 10^{19}$&$2.01 \cdot 10^{19}$\tabularnewline\hline \end{tabular}
\end{center}
\caption[]{\label{tab:etajjtildehigh}Same as \fref{tab:etajjhigh}, but for a
  pseudo-scalar Higgs (see \fref{fig:etadist-pseudo-high-high}).}
\end{table}

\end{appendix}


\clearpage

\def\app#1#2#3{{\it Act.~Phys.~Pol.~}\jref{\bf B #1}{#2}{#3}}
\def\apa#1#2#3{{\it Act.~Phys.~Austr.~}\jref{\bf#1}{#2}{#3}}
\def\annphys#1#2#3{{\it Ann.~Phys.~}\jref{\bf #1}{#2}{#3}}
\def\cmp#1#2#3{{\it Comm.~Math.~Phys.~}\jref{\bf #1}{#2}{#3}}
\def\cpc#1#2#3{{\it Comp.~Phys.~Commun.~}\jref{\bf #1}{#2}{#3}}
\def\epjc#1#2#3{{\it Eur.\ Phys.\ J.\ }\jref{\bf C #1}{#2}{#3}}
\def\fortp#1#2#3{{\it Fortschr.~Phys.~}\jref{\bf#1}{#2}{#3}}
\def\ijmpc#1#2#3{{\it Int.~J.~Mod.~Phys.~}\jref{\bf C #1}{#2}{#3}}
\def\ijmpa#1#2#3{{\it Int.~J.~Mod.~Phys.~}\jref{\bf A #1}{#2}{#3}}
\def\jcp#1#2#3{{\it J.~Comp.~Phys.~}\jref{\bf #1}{#2}{#3}}
\def\jetp#1#2#3{{\it JETP~Lett.~}\jref{\bf #1}{#2}{#3}}
\def\jphysg#1#2#3{{\small\it J.~Phys.~G~}\jref{\bf #1}{#2}{#3}}
\def\jhep#1#2#3{{\small\it JHEP~}\jref{\bf #1}{#2}{#3}}
\def\mpl#1#2#3{{\it Mod.~Phys.~Lett.~}\jref{\bf A #1}{#2}{#3}}
\def\nima#1#2#3{{\it Nucl.~Inst.~Meth.~}\jref{\bf A #1}{#2}{#3}}
\def\npb#1#2#3{{\it Nucl.~Phys.~}\jref{\bf B #1}{#2}{#3}}
\def\nca#1#2#3{{\it Nuovo~Cim.~}\jref{\bf #1A}{#2}{#3}}
\def\plb#1#2#3{{\it Phys.~Lett.~}\jref{\bf B #1}{#2}{#3}}
\def\prc#1#2#3{{\it Phys.~Reports }\jref{\bf #1}{#2}{#3}}
\def\prd#1#2#3{{\it Phys.~Rev.~}\jref{\bf D #1}{#2}{#3}}
\def\pR#1#2#3{{\it Phys.~Rev.~}\jref{\bf #1}{#2}{#3}}
\def\prl#1#2#3{{\it Phys.~Rev.~Lett.~}\jref{\bf #1}{#2}{#3}}
\def\pr#1#2#3{{\it Phys.~Reports }\jref{\bf #1}{#2}{#3}}
\def\ptp#1#2#3{{\it Prog.~Theor.~Phys.~}\jref{\bf #1}{#2}{#3}}
\def\ppnp#1#2#3{{\it Prog.~Part.~Nucl.~Phys.~}\jref{\bf #1}{#2}{#3}}
\def\rmp#1#2#3{{\it Rev.~Mod.~Phys.~}\jref{\bf #1}{#2}{#3}}
\def\sovnp#1#2#3{{\it Sov.~J.~Nucl.~Phys.~}\jref{\bf #1}{#2}{#3}}
\def\sovus#1#2#3{{\it Sov.~Phys.~Usp.~}\jref{\bf #1}{#2}{#3}}
\def\tmf#1#2#3{{\it Teor.~Mat.~Fiz.~}\jref{\bf #1}{#2}{#3}}
\def\tmp#1#2#3{{\it Theor.~Math.~Phys.~}\jref{\bf #1}{#2}{#3}}
\def\yadfiz#1#2#3{{\it Yad.~Fiz.~}\jref{\bf #1}{#2}{#3}}
\def\zpc#1#2#3{{\it Z.~Phys.~}\jref{\bf C #1}{#2}{#3}}
\def\ibid#1#2#3{{ibid.~}\jref{\bf #1}{#2}{#3}}
\def\otherjournal#1#2#3#4{{\it #1}\jref{\bf #2}{#3}{#4}}
\newcommand{\jref}[3]{{\bf #1} (#2) #3}
\newcommand{\hepph}[1]{{\tt [hep-ph/#1]}}
\newcommand{\mathph}[1]{{\tt [math-ph/#1]}}
\newcommand{\arxiv}[2]{{\tt arXiv:#1}}
\newcommand{\bibentry}[4]{#1, {\it #2}, #3\ifthenelse{\equal{#4}{}}{}{, }#4.}


\end{document}